\documentclass[a4paper,11pt]{article}

\usepackage{fullpage,amsmath,amssymb,enumerate,color}
\usepackage[pdftex]{graphicx,psfrag}
\usepackage{wrapfig}
\usepackage[font={small}]{caption}
\usepackage{subfig}
\usepackage{listings}
\usepackage{varioref}

\usepackage{relsize}
\usepackage{authblk}
\usepackage{gensymb}

\providecommand{\keywords}[1]{\textbf{\textit{Keywords---}} #1}
\providecommand{\abbreviations}[1]{\textbf{\textit{Abbreviations---}} #1}
\usepackage{hyperref}

\makeatletter
\newcommand{\vast}{\bBigg@{3.5}}
\newcommand{\Vast}{\bBigg@{4}}
\makeatletter \renewcommand{\fnum@figure}
{\figurename~S\thefigure}
\makeatother

\def\be{\begin{equation}}
\def\ee{\end{equation}}
\def\bea{\begin{eqnarray}}
\def\eea{\end{eqnarray}}

\title{Nonequilibrium mechanisms underlying de novo biogenesis of Golgi cisternae}
\date{}
\author[1,2]{Himani Sachdeva}
\author[1,3]{Mustansir Barma}
\author[4,5]{Madan Rao*}
\affil[1]{Department of Theoretical Physics, Tata Institute of Fundamental Research, Homi Bhabha Road, Mumbai 400005, India}
\affil[2]{Institute of Science and Technology, Am Campus 1, Klosterneuburg A-3400, Austria}
\affil[3]{TIFR Centre for Interdisciplinary Sciences, 21 Brundavan Colony, Narsingi, Hyderabad 500075, India}
\affil[4]{Raman Research Institute, C.V. Raman Avenue, Bangalore 560080, India}
\affil[5]{Simons Centre for the Study of Living Machines, National Centre for Biological Sciences (TIFR), Bellary Road, Bangalore 560065, India}

\begin{document}

{\maketitle
\begin{abstract}
A central  issue in cell biology is the physico-chemical basis of organelle biogenesis in intracellular trafficking pathways, 
its most impressive manifestation being the biogenesis of Golgi
cisternae. At a basic level, such morphologically and chemically distinct compartments should arise from an
 interplay between the molecular transport and chemical maturation.
Here, we formulate
 analytically tractable, minimalist models,
 that incorporate this interplay between transport and chemical progression in physical space, and
  explore the conditions for {\it de novo} biogenesis of distinct cisternae.
  We propose new quantitative measures that can discriminate between the various models of transport in a {\it qualitative} manner - 
  this includes measures of the dynamics in steady state and the dynamical response to perturbations of the kind amenable to live-cell imaging.

 \end{abstract}}


\abbreviations{MC, Monte Carlo; VT, vesicular transport; CP, cisternal progression}

  
\paragraph*{}

One of the striking features of eukaryotic cells, 
is the appearance of compartmentalization,
especially  under continual remodeling,
 such as in  the endosomal or secretory trafficking pathways \cite{alberts_molbio_ch5, Diekmann_2013}.
This naturally raises the question,
 are there robust self-organizational principles that lead to
 the emergence of  chemically and morphologically distinct compartments in such dynamic situations  \cite{Misteli_2001, glickmalhotra1998}?




Such questions can be discussed in the context of the Golgi Apparatus, which consists of
multiple stacks of chemically distinct, membrane-bound cisternae \cite{Polishchuk}.
Starting from their point of synthesis in the endoplasmic reticulum (ER), lipids and proteins are 
 transported vectorially through the 
 polarized
 Golgi stacks towards the plasma membrane (PM), undergoing a sequence of enzymatic conversions during their progression. 
 This interplay between vectorial transport and
  biochemical polarity appears quite generally across different cell types, though
  the number, shape and spatial extent of the Golgi cisternae may exhibit variation \cite{GlickNakano2009}.


A detailed knowledge of {\it individual molecular processes} 
notwithstanding,
 a step in understanding the {\it collective dynamics} of morphological and chemical identity is to construct minimalist 
 theoretical models \cite{Heinrich,gong2008,kuhnle2010,Ispolatov,sensrao2013,thattai} that take into account the essential microscopic dynamical and chemical processes in the Golgi,
  to arrive at general conditions for compartment biogenesis.
Such an exercise would be useful in addressing the 
 fundamental issue of whether the Golgi organelle is
 constructed from a pre-existing template  or generated {\it de novo}, a result of self-organization \cite{glickmalhotra1998,GlickNakano2009, PuriLinstedt2003,Lowe_2007}.
 In the process, it may lead us to revisit and make precise, the notion of an organelle.

Our minimalist theoretical framework incorporates the two classical competing models for intra-Golgi transport,
(A) \emph{Vesicular Transport (VT)} and (B) \emph{Cisternal Progression  and Maturation (CP)}. In the VT model, cisternae are assumed to be stationary and temporally stable structures, each with 
 a fixed set of enzymes. Molecules, packaged within vesicles, shuttle from one cisterna to the next, and get 
chemically modified by the resident enzymes.
By contrast, in the CP model, there is a constant turnover of cisternae-- new cisternae form by the fusion of 
incoming vesicles at the cis end, and move \emph{as a whole} through the Golgi, carrying cargo molecules with them. Specific enzymes get attached to a cisterna in different stages of its progression, and 
modify its contents. The terminal (trans) cisterna is the site of extensive tubulation and budding, whereby 
processed biomolecules are released onto the PM \cite{alberts_molbio_ch5, Glick_models}.



Experimental evidence in support of the two hypotheses, namely, VT  and CP \cite{bonfati1998,cisternalmat_glick,cisternalmat_nakano}
or their combinations \cite{Marsh_2001_mixedtransport, Beznoussenko,pelham_ch5}, is mixed at best. To our knowledge, except in the few cases, 
it has been difficult to implement an experimental strategy 
 that can point unequivocally to a specific transport model. Indeed
natural cellular realizations might have aspects of both these models.
One of our objectives is to suggest new quantitative measures that can discriminate between various contending models.
In order to do this,  we construct a general multi-species model that encompasses essential features of both VT and CP and their many variants (such as
the recently proposed `cisternal progenitor' and `rim progression' models \cite{Pfeffer, Lavieu} ). Our generic framework allows us to take each of these 
models and explore its consequences in detail. 

%
%


The rules governing the model are kept 
fairly general, with no assumptions of molecular constraints or selectivity, hoping to arrive at a good 
compromise between complexity and analytical tractability.
The general model can be analyzed precisely in various limits, to determine the conditions under which one might obtain multiple 
compartments that are 
large, clearly separated, chemically distinct, 
statistically stable and robust to perturbations - properties that are 
 desirable for organelle biogenesis.
Further, we propose precise biophysical measurements,  associated with subtle features of the dynamics in steady state and dynamical response to 
 a variety of perturbations,
 that show 
{\it qualitatively} distinct signatures 
of the underlying transport mechanism.

\begin{figure*}
 \centering
 \includegraphics[width=0.6\textwidth]{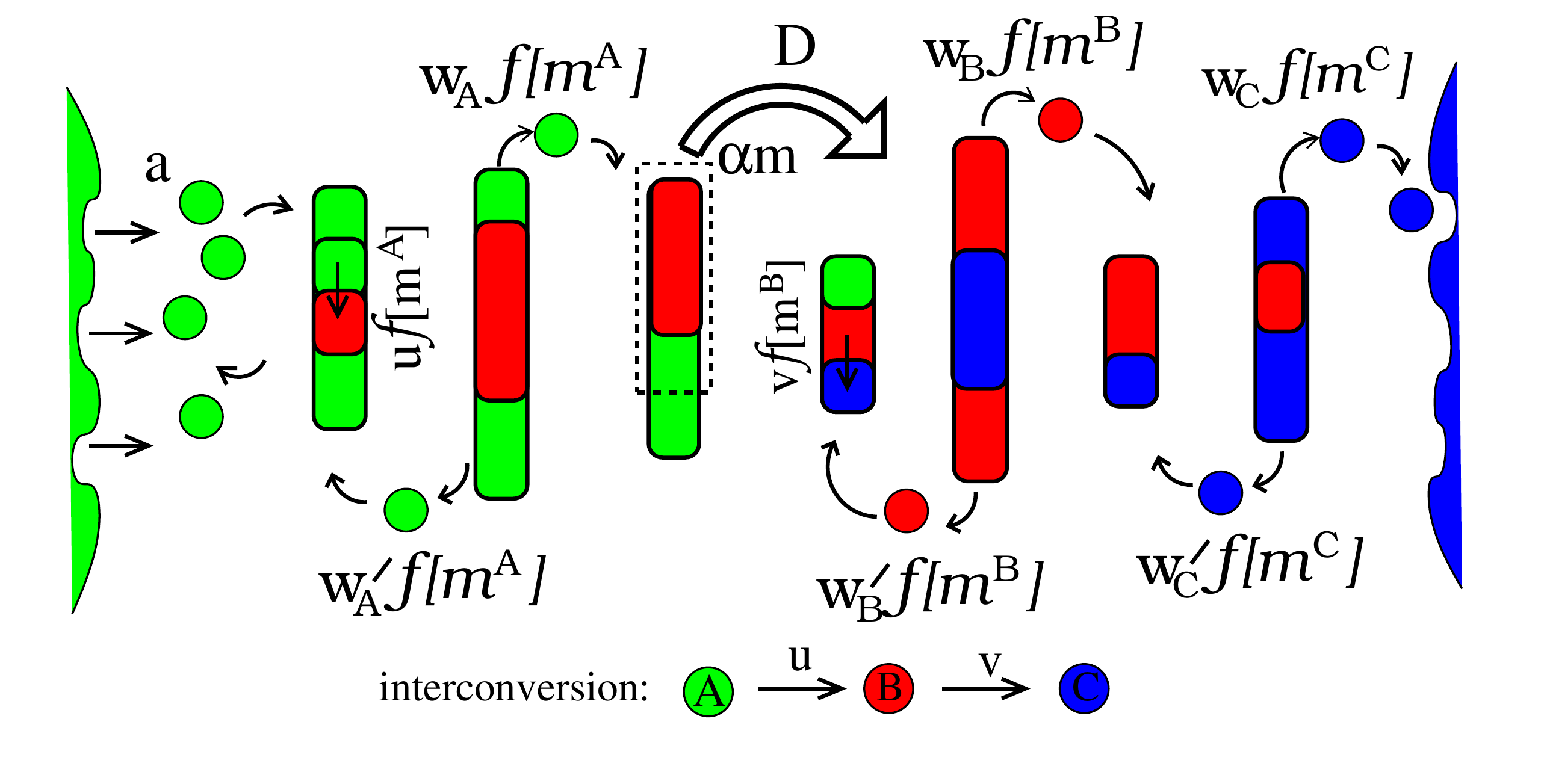}
 \caption{Schematic of the effective 1D transport model with three-species -- $A$, $B$ and $C$ of particles (vesicles). 
 The stochastic model
 includes injection from the ER (left) with rate $a$, fission-fusion (with rates $w_{A}, w_{B}, w_{C}, \ldots; w^{\prime}_{A}, w^{\prime}_{B}, w^{\prime}_{C},\ldots$) of vesicles, aggregate breakage (fraction $\alpha$) and movement (rate $D$), 
 chemical interconversion (rates $u, v, \ldots$) and  exit from the PM (right boundary) or recycling back to the ER (left boundary). The rates of transport, fission, interconversion 
 depend on the amount or ``mass'' of the chemical species at the donor and acceptor sites through the flux-kernel $f$.
 Rates of various processes depicted in the figure are detailed in {\it Supplementary Information}.
 }
 \label{fig1m}
 \end{figure*}

\section*{Model and Methods}

Our general model is defined in terms of 
dynamical moves of  ``particles'' (vesicles) each carrying a chemical species, denoted as $A, B, C, \ldots$, in a 1-dimensional (1D) system--
for convenience represented  as
a 1D lattice 
with $L$ sites. This could describe  the endosomal and secretory pathways where the primary transport, barring a few branching events, is along 1D.
However for definiteness, we will henceforth use the terminology of the secretory pathway.
The model allows for (i) influx of unprocessed cargo at the 
left (ER) in the form of injection of A particles, (ii) outflux of processed molecules at the right (PM), i.e., exit of particles from the right boundary, (iii) recycling of particles back to the ER, i.e., 
exit of particles from the left, 
(iv) fusion of particles to form an aggregate and fusion to a preexisting aggregate, (v)
fission from an aggregate, (vi) transport within the system either via single particle 
movement (vesicle exchange) and/or collective movement of the entire aggregate of particles (cisternal movement), and finally, 
(vii) chemical transformation or processing of molecules in the aggregates, via the sequential interconversion $A\rightarrow  B \rightarrow C \rightarrow \ldots$
of particles. For further details, see {\it Supplementary Information} (also schematic in Fig.\,\ref{fig1m}).

Directional transport  is modeled via asymmetric anterograde/retrograde particle movement rates, 
with $w_{A}\rightarrow \gamma_{A} w_{A}$, $w^{\prime}_{A}\rightarrow \left(1-\gamma_{A}\right) w_{A}$, where $\gamma_{A}$ parametrizes the asymmetry for A particles and so on (Fig.\,\ref{fig1m}). This asymmetry factor could depend on the activity of GTPases such as CDC42 \cite{cdc42}.
%
The dependence of the asymmetry factors $\gamma_{A}$, $\gamma_{B}$, $\gamma_{C}$ on the chemical species
corresponds to differential degrees of 
recycling of $A, B, C, \ldots$ particles to the ER.

%

The rates of fission-fusion and chemical conversion depend on the amount (or ``mass'') of chemical species in 
the donor ($m_i$) and acceptor ($m_{i\pm1}$) aggregates, through the flux-kernel, $f[m_i \vert m_{i\pm1}]$.
 There are many choices of the flux-kernel (see {\it SI} for a discussion), but the form 
that we explore in detail is independent of the mass  of the acceptor aggregate:
 \begin{equation}
 f[m_i \vert m_{i\pm1}]=\frac{K_{sat}m_i^{\theta}}{m_i^{\theta}+m^{\theta}_{sat}}, \,\,\,\,\,\, 0 \leq \theta \leq 1
 \label{eq1}
 \end{equation}
and corresponds to Michelis-Menten (MM) type of enzymatic kinetics
 for fission or chemical conversion, where we choose $\theta = 1/2$ \footnote{
 Our choice of  $\theta=1/2$
is based on the realization that our effective 1D model is obtained by integrating over the two 
directions perpendicular to the cis to trans axis in the Golgi,  and
assuming that only the perimeter molecules participate in the flux.
}.
 Thus, reaction rates increase as $\sqrt{m^{A}}$, $\sqrt{m^{B}}, ..$ for small amounts of $A$, $B$ in the parent aggregate, but become enzyme-limited and saturate to a constant value $K_{sat}$ when the 
 amount of $A$, $B$ exceeds $m_{sat}$.
In principle, the saturation scale $m_{sat}$ can be different for fission and conversion 
processes, but to keep the model analytically tractable 
(using techniques developed in  \cite{Himani2011m}), we restrict it to be the same for both. 
 We have also studied flux-kernels which
depend on the amount of the chemical species in the acceptor aggregate, and find that this does does not qualitatively change the results (see SI).



Note that chemical interconversion in our model is a highly simplified representation of the sequential processing of 
proteins as they pass through the Golgi  \cite{Stanley_enzymes}. Instead of explicitly incorporating the enzymes responsible for various biochemical modifications, 
we treat these modifications as simple Poisson processes occurring with specified rates, as in 
\cite{Sens}. The rates should thus be viewed as effective or composite parameters.
Despite its simplicity, we believe that this qualitatively captures the role of the interconversion process in shaping the macro organization of the Golgi.


The transport and chemical interconversion rules described above, can also be re-expressed as dynamical equations for the average ``mass'' of each chemical 
species at each site.
 Under suitable conditions, these equations can be solved to obtain the spatial
profiles of the mass (abundance) of each species in steady state. From these profiles will emerge features such as `compartments', for which we provide a precise
definition below.

\subsection*{Definition of a compartment}
Given the finite spatiotemporal resolution of coarse-grained models, 
one needs to provide a consistent definition of a compartment, an aspect that is  also pertinent for in-vivo
imaging of Golgi compartments.
As we will see, this is particularly germane to the VT model and its variants. 

\begin{figure*}[h!]
\centering
   \renewcommand{\thesubfigure}{a}
\subfloat[]{\label{fig2a1}
  \includegraphics[width=7.3cm,height=5.0cm]{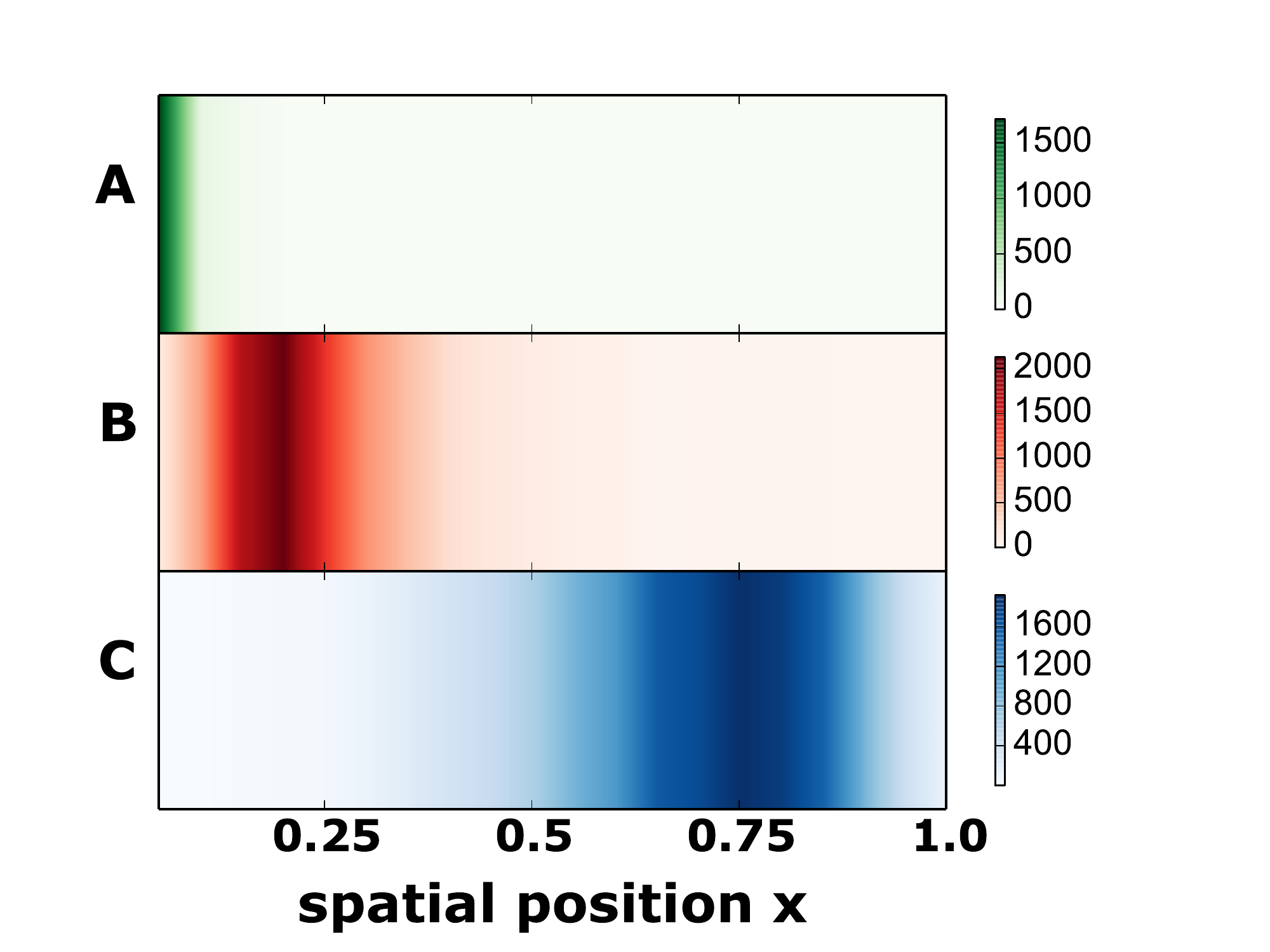}
  }
  \renewcommand{\thesubfigure}{b}
\subfloat[]{\label{fig2b1}
  \includegraphics[width=7.3cm,height=5.0cm]{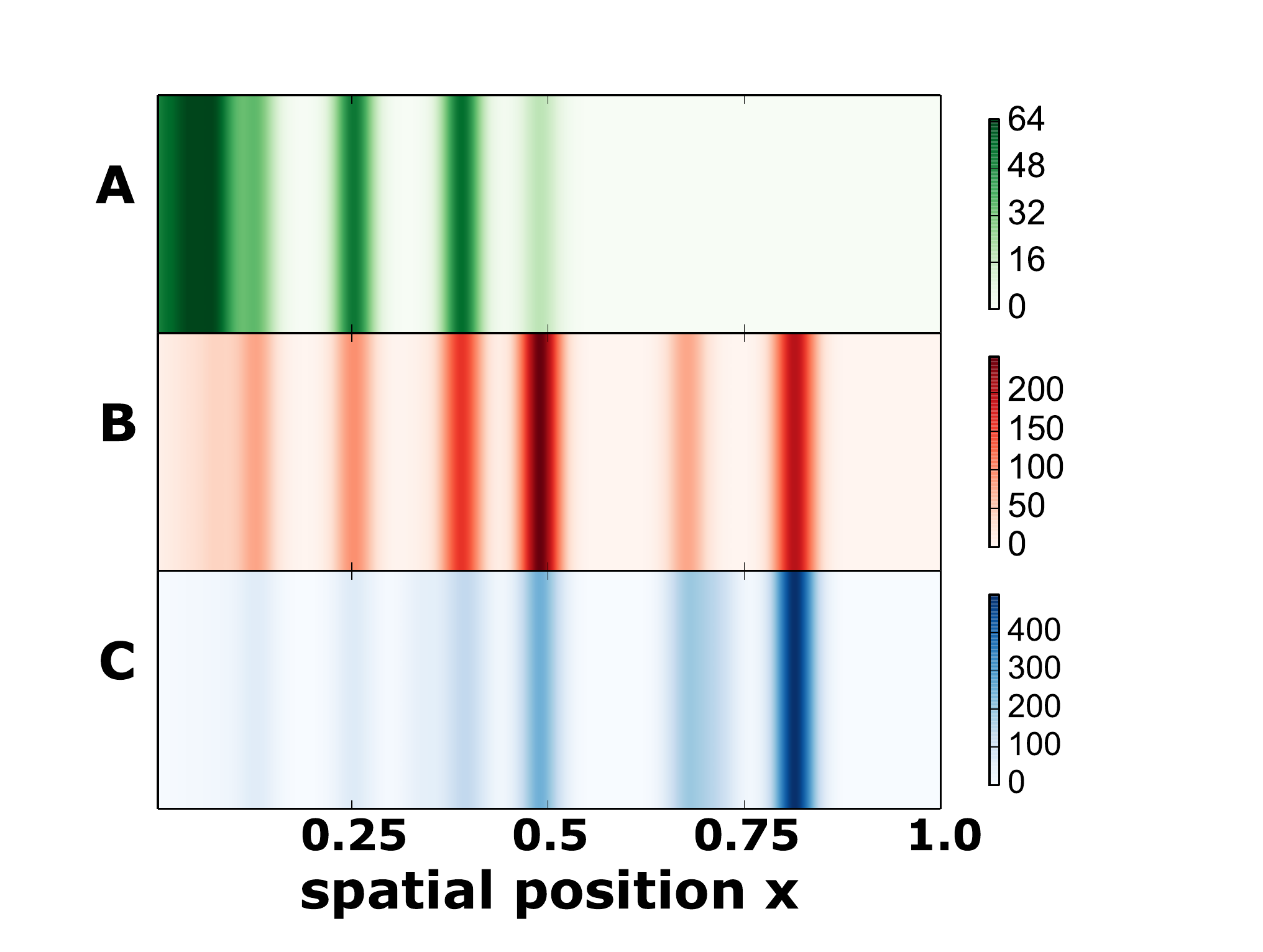}      
  }  
  \\
 \renewcommand{\thesubfigure}{c}
  \subfloat[]{\label{fig3a1}
  \includegraphics[width=7.3cm,height=5.0cm]{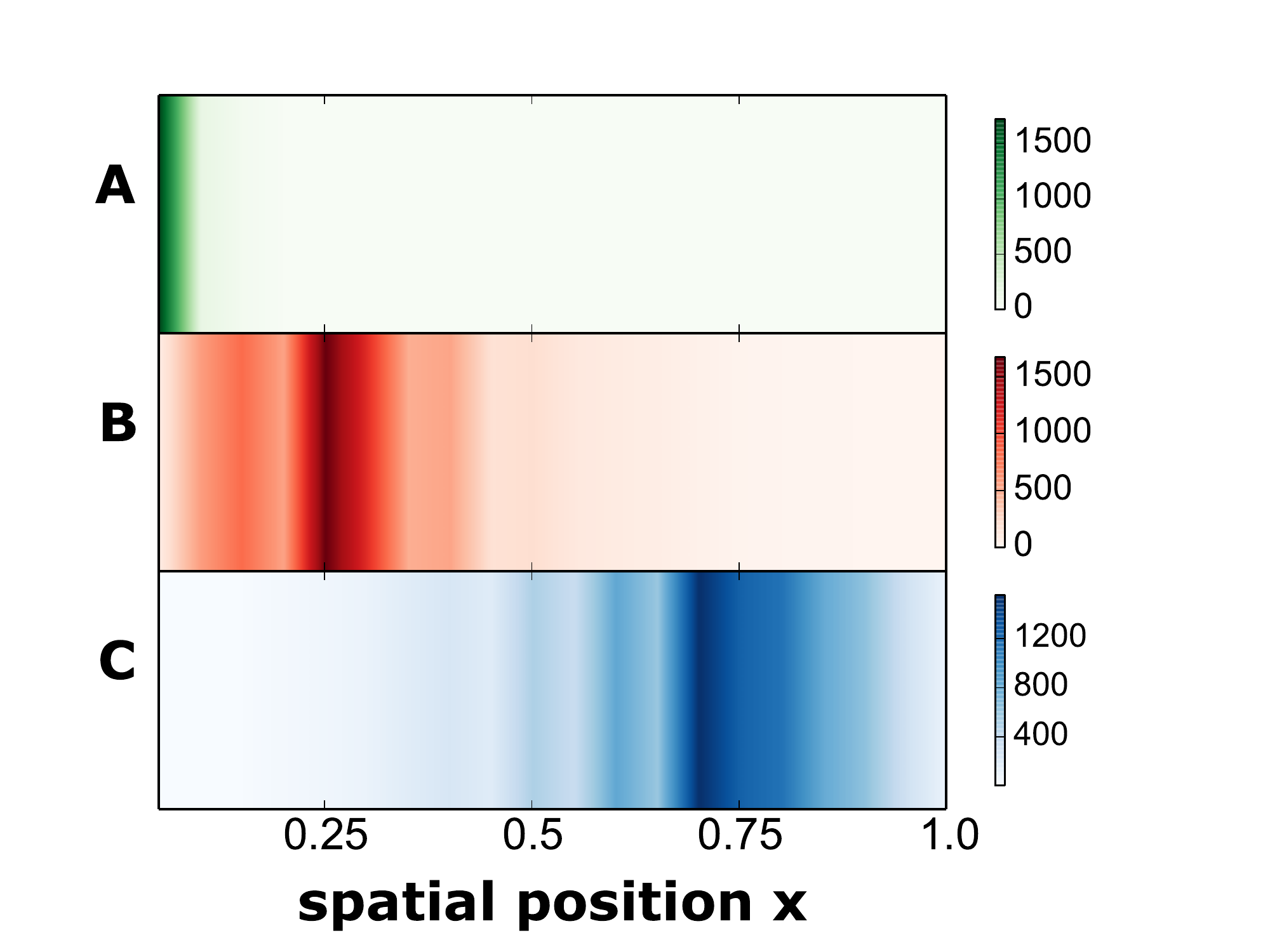}
  }
  \renewcommand{\thesubfigure}{d}
\subfloat[]{\label{fig3b1}
  \includegraphics[width=7.3cm,height=5.0cm]{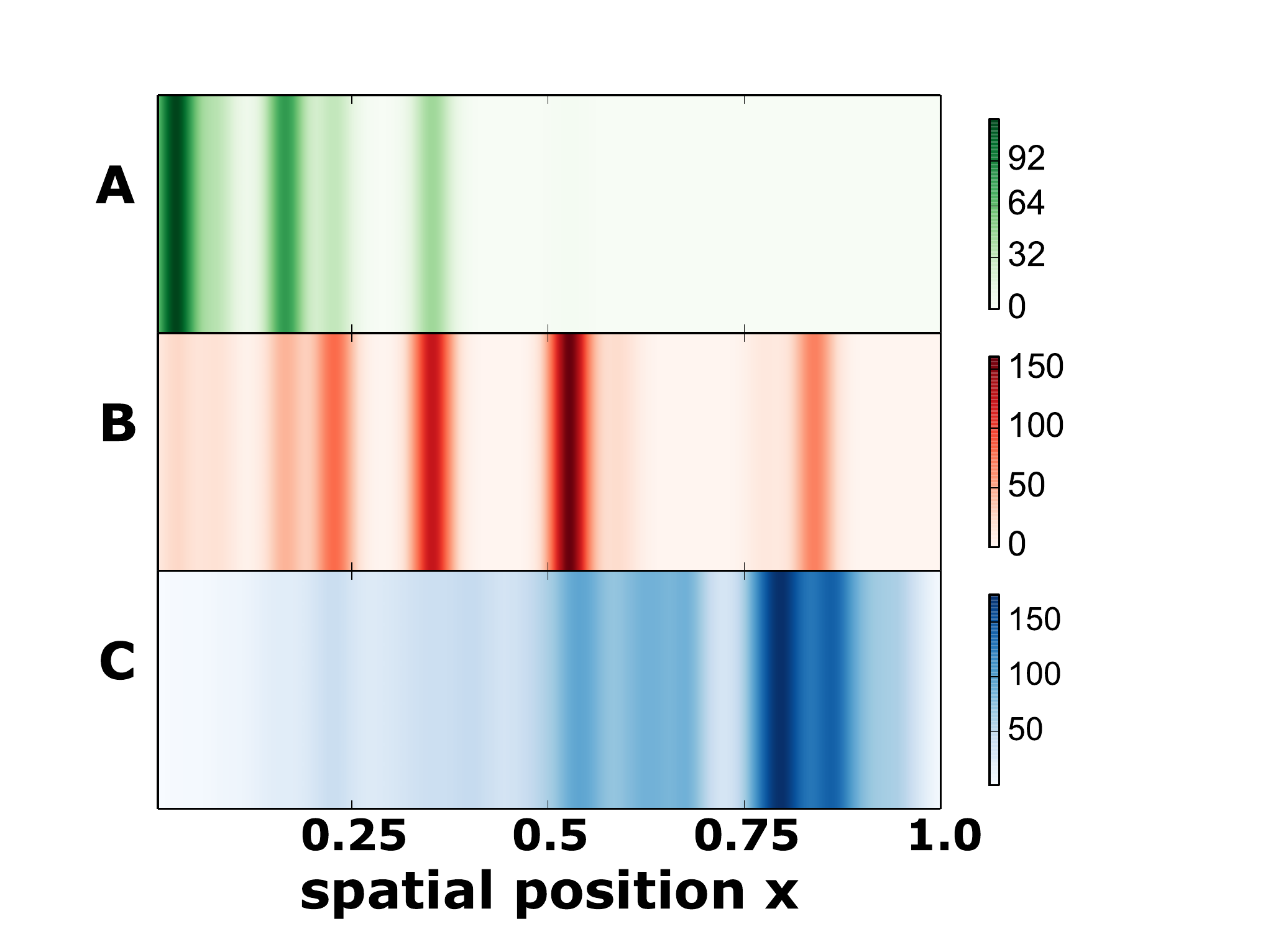}
  }
\caption{Typical steady state configurations for (a) pure VT (b) pure CP (c) VT-dominated transport with rare sub-cisternal movement, akin to the Cisternal Progenitor model and 
(d) CP-dominated transport with fission of single C particles.
All configurations depicted
in Gaussian-smoothed {\it compartment representation} where color intensity in 
each species channel indicates species mass (see LUT bar). 
(a) Pure VT model  shows well-separated, pure (single-color) compartments while (b) pure CP model shows many mixed-color compartments with a gentle cis to trans gradient of each species.
(c)  VT-dominated transport: Mass profiles are 
qualitatively similar to the pure VT case, but with higher number of cisternal fragments between compartments.
(d) CP-dominated transport: Sharp aggregates similar to those
in the CP model are found in cis and medial regions of system, but disintegrate into smaller aggregates near the trans end.
%
 Parameters for (a): $a=1$, $D=0$, $w_{A}=0.125$, $w_{B}=0.04375$, $w_{C}=0.05462$, $\gamma_{A}=0.5$, $\gamma_{B}=0.6$, 
 $\gamma_{C}=0.66$, $u=0.01125$, $v=0.00194$, $m_{sat}=200$, $K_{sat}=14.14$, $L=20$;
(b): $a=1$, $D=0.00833$, $w_{A}=w_{B}=w_{C}=0$, $u=0.025$, $v=0.01166$, $m_{sat}=200$, $K_{sat}=14.14$, $L=80$.    
 (c): $a=1$, $D=0.000125$, $\alpha=0.3$, $w_{A}=0.125$, $w_{B}=0.0425$, $w_{C}=0.05375$, $\gamma_{A}=0.5$, $\gamma_{B}=0.6$, 
$\gamma_{C}=0.66$ $u=0.01125$, $v=0.001937$, $m_{sat}=200$, $K_{sat}=14.14$, $L=20$; 
(d): $a=1$, $D=0.00833$, $w_{A}=w_{B}=0$, $w_{C}=0.8333$, $u=0.025$, $v=0.01662$, $m_{sat}=200$, $K_{sat}=14.14$, $L=80$.    
}
  \label{fig2m}
\end{figure*}

To identify self-organized compartments we locate the maxima (concentration peaks) in the spatial profile of the total mass and note the locations on either side of the maximum where the mass drops to half its peak value. This half-width around maximum 
is taken to be the compartment size (and can span many lattice sites) while  the sites between adjacent compartments constitute intercisternal regions. While defining dynamical moves (Fig.\,\ref{fig1m}),
however, we represent  cisternal or sub-cisternal progression as the collective movement of the mass at a single site,  rather than the movement of this many-site agglomerate. This leads
to a consistent description 
 in the CP limit where a compartment has a
small width (spanning 1-2 sites). In the VT limit, where a typical compartment can span many sites (see Fig.\,\ref{fig2a1}), the breakage, movement and fusion of sub-cisternal fragments from site to site within the compartment can
be viewed as an intra-compartment remodeling process which  eventually leads to 
the budding of a large fragment from the anterograde face of the compartment and its fusion with the next compartment.
\subsection*{Units}
Distances are shown in terms of the position variable $x$, which is scaled by $L$, the distance between the cis and trans end. Masses are defined as multiples of the unit mass, which is the 
typical mass of a vesicle or `particle' breaking off from a cisterna (see Fig.\,\ref{fig1m}). A time unit (t.u.) is chosen such that the injection rate $a$ is $1$ vesicle/t.u..

\section*{Results}
\section*{Structure formation in steady state}
We now demonstrate how the interplay of various dynamical processes in the model gives rise to
 steady states configurations that mirror the spatial organization of the Golgi cisternae. 
We first consider the two extreme limits of the model--
the \emph{pure VT} and the \emph{pure CP} limits,
%
  and then explore variants around these 
  limiting cases, with the aim of modeling `mixed' traffic strategies.
\subsection*{Structure formation with  pure vesicular transport}
\hspace{1cm} The pure vesicular transport (VT) model is obtained
by setting
the aggregate movement rate $D=0$.
This allows us to solve for the average mass of each species at each site  
in steady state, using the techniques described in  \cite{Himani2011m} (see SI for details).
While we have explored several functional forms of $f[m]$ using numerical simulations (SI), we discuss below 
properties of structures generated using Eq.\,\ref{eq1}.

An example of  
 compartment emergence in the pure VT limit for the three-species model with $A\rightarrow B\rightarrow C$ 
conversion, is depicted in Fig.\,\ref{fig2a1}.
A typical snapshot shows that while $A$ particles accumulate close to the cis end, $B$ and $C$ particles form distinct compartments in the interior. 
Similarly, the four-species model exhibits steady states with  A particles near the cis end and 
 distinct B, C and D structures in the interior (see SI).
Self-organization of compartments in the pure VT limit is driven by three main effects, which we discuss below.
\paragraph*{A.}
{\bf Spatial Organization of Compartments}. The spatial segregation of chemical species and their localization in different regions, depends on  
(i) the ratio of chemical conversion rate to fission rates, $u/w_A, \ldots $, and (ii) the degree of directional bias in transport. 

If the ratio $u/w_{A}$ is very large, most A particles undergo an $A\rightarrow B$ conversion before they can travel into the interior,
leading to an accumulation of B particles close to the source itself. Conversely, if $u/w_{A}\ll 1$, B particles localize 
  away from the source. The length scale governing the localization of 
C particles depends likewise on the ratio $v/w_{B}$. The second  determinant of the region of localization is the  
 directional bias $\gamma_{A}$, $\gamma_{B}$, $\gamma_{C}$
 -- particles with greater recycling to the source localize closer to the cis end, while those with less 
 recycling form structures near the trans end.
Thus, if the directional bias factors $\gamma_{B}$ and $\gamma_{C}$ are unequal, and the 
ratios $u/w_{A}$ and $v/w_{B}$ sufficiently different, the peak concentration of $B,C, \ldots$ occur at well-separated locations. 
 \paragraph*{B.} {\bf Morphologically Distinct Compartments}.
Spatially localized particles of a particular species form a large, morphologically distinct 
aggregate only if the flux-kernel $f[m]$ 
attains  a constant value 
at large $m$ (as in Eq.\,\ref{eq1}),  i.e.,
the 
outflux of vesicles from any aggregate saturates as aggregate size increases. 

  Sharp localization of B, C particles in different regions 
(as opposed to gentle gradients) is the key to obtaining spatially resolvable cisternae-like structures. To see how 
 flux-kernels that saturate at large $m$ lead to sharp localization, consider how the structures in  Fig.\,\ref{fig2a1} respond to a 
 small increase in influx. An increase in influx increases the particle concentration or mass $m$ in all regions of the system, which in turn causes the vesicular outflux (proportional to $f[m]$) to rise until 
 a new flux balance is achieved. However, in regions of peak concentration, where $m\gg m_{sat}$ and the vesicular outflux $f[m]$ is already in the saturation regime,
the mass has to increase substantially in order to achieve the small increase in vesicular outflux required for steady state. Conversely, in the intercisternal regions, where $m$ is small and $f[m]$ is not in the 
saturation regime, a modest increase in mass is enough to achieve the requisite increase in vesicular outflux. Thus, with saturating flux kernels $f[m]$ of the type in Eq.\,\ref{eq1}, the contrast in particle concentration between 
cisternal and intercisternal regions is naturally amplified when influx is high, leading to the emergence of sharply localized and well-defined compartments
In fact, if the influx is too high, 
the vesicular outflux may fail to balance the influx near the peaks, leading to runaway 
growth of mass, as elaborated in \cite{Himani2011m} (also pointed out in \cite{Sens}).  
The key to generating large but finite concentrations of A,B,C  particles in specific regions, is to tune the  
 influx rate $a$ such that the system is in steady state, but 
{\it poised close to the transition point} to runaway growth. This causes the peaks in the average mass profile to become very sharp.

\paragraph*{C.}
{\bf Robustness of Compartments}.
There is a trade-off between the temporal stability of a compartment and how sharply localized it is; 
this can be optimized with a flux-kernel $f[m]$ that increases for small $m$ but saturates at large $m$ (as in Eq.\,\ref{eq1}).

Stability of structures can be quantified using the ratio of the root mean square
 (rms) fluctuations $\Delta m$ to the mean mass $\langle m\rangle$ at any location.
We measure this ratio for different functional forms of 
$f[m]$ in our simulations (details in SI), and find that  aggregates are stable 
($\Delta m/\langle m\rangle\rightarrow 0$ for large $\langle m\rangle$),  if the fission rate and hence the flux-kernel 
$f[m]$ increases with $m$. This ensures a stabilizing negative feedback-- when
 the aggregate becomes too large, the number of particles breaking off from it increases and vice versa, thus 
 restoring the aggregate to its average size. 
However, as discussed above, sharply localized A,B,C compartments emerge only if vesicular fluxes are 
 relatively insensitive to aggregate size, i.e., for $f[m]$ that saturates at large $m$.
The best compromise between the competing requirements of stability and 
sharp localization is obtained with MM (or qualitatively similar) fission rates, as in Eq.\,\ref{eq1}.

Thus,  spatially and chemically distinct compartments  
emerge in the pure VT limit, even in the \emph{absence of a pre-existing template} if the three conditions above are satisfied
(for a more rigorous discussion of the conditions, see SI). 
A limitation of this mode of self-organization is that it requires {\it fine tuning} of parameters, and that 
structures may even undergo an instability, leading to runaway growth,  in certain regions of 
parameter space. However, this instability is  suppressed if we allow for a mixed transport strategy, in which 
vesicular transport is accompanied by rare  sub-cisternal movement (see below). 
 \subsection*{Structure formation with pure cisternal progression}
 The cisternal progression and maturation (CP) limit, obtained by setting $w_{A}=w_{B}=w_{C}=0$ and  $\alpha=1$, 
 allows for anterograde movement of full aggregates at rate $D$
(with aggregates fusing on encounter to form larger aggregates),
in addition to the sequential $A\rightarrow B\rightarrow C\rightarrow$ conversion of particles 
within each aggregate.

This system self-organizes into a state where the entire mass is concentrated in a relatively small number ($\sim\sqrt{L}$) of aggregates (see Figs.\,\ref{fig2b1}).
The  typical mass of an aggregate increases with increasing distance $x$ from the (ER) source,
while the probability of finding an aggregate decreases with $x$ \cite{Himani2013}.  Thus, compartments become larger but 
sparser from cis to trans. 
The emergence of morphologically separate structures is a generic feature of this model, and occurs {\it without any fine tuning}, because of the
tendency of large moving aggregates to sweep up all the mass in their vicinity.
The sequential conversions further cause the system to develop biochemical polarity, 
with the composition of aggregates showing a gentle gradation from 
 A-rich to B-rich to C-rich in the cis to trans direction.
The length scales governing these chemical gradients depend
on the interconversion rates and the function $f[m]$. System-wide gradients emerge for 
$u, v \ll D$, i.e., for conversion rates that are much smaller than the rate of cisternal movement. 
A distinguishing feature of structure formation in the CP limit is the occurrence of a large number of mixed (two-color) compartments (Figs.\,\ref{fig2b1}).

The movement of large aggregates through the system also results in distinctive
dynamical properties: the total mass in the system undergoes \emph{intermittent} time evolution \cite{Himani2013}. 
We discuss later how this can be a revealing signature of cisternal passage.
Most of these qualitative features of structure formation by CP appear robust upon changing transport rules, e.g., by making the 
aggregate movement rate $D$ mass or chemistry dependent.

\subsection*{Comparing structure formation in the two limits}
While multiple, 
spatially and chemically distinct compartments emerge in both pure VT and CP models, there are significant differences between structures in the two limits that we discuss below.

First, structures generated in the pure VT model (with MM rates) are \emph{stable} (position, size and composition vary
little over time), though highly sensitive to changes in the rates of underlying processes.
By contrast,
 in the pure CP model, structures are highly \emph{dynamic}, with
their positions, composition and even number showing significant variation over time. 
Monitoring the extent of variability 
of features such as the number of cisternae and the fluctuations in fluorescence from specific Golgi regions 
 over time (see next section) using live-cell imaging
  could thus  provide valuable clues to the mechanism of transport (to a limited extent this has been done in \cite{Mukherji2014}).

 Second, while
 interconnected cisternae are observed in the VT limit (Fig.\ref{fig2a1}, also figs. 4(e)-4(f) of SI), 
 whereas more discrete cisternae (with little or no mass in the intercisternal regions) arise in the CP limit
(Fig.\ref{fig2b1}). While the exact nature of this prediction
may hinge on specifics of the model, the observation points towards an interesting correlation between the presence of 
inter-cisternal tubulation and the mode of intra-Golgi traffic.


Finally, the two limits also differ in
the degree of spatial localization  of individual chemical species. In the
VT model, each species is \emph{sharply localized} and not spread over multiple compartments
(Figs. \ref{fig2a1}).
%
In the CP model, a typical snapshot
 (Figs.  \ref{fig2b1}) reveals   
 many aggregates of mixed composition, with the relative abundance of each species  exhibiting a gentle gradient across aggregates in the cis to trans direction.
 This distinction is fairly robust, suggesting possible experimental strategies aimed at exploring the
 connection between transport mechanisms and the spatial localization/delocalization of enzymes and markers, a connection that was also highlighted in \cite{Opat_2001}.

\subsection*{Structure formation with mixed transport strategies}
We now consider `mixed' strategies of transport, which involve features of both the VT and the CP models to different 
extents
\cite{pelham_ch5, dmitrieff}. Since the parameter space is very large, we study only small deviations
from either limit. 
This is also important from a stability  
perspective, since a mechanism for maintaining differentiated compartments must be robust to small variations in the model. 

\paragraph*{Cisternal progenitor as a small deviation about pure vesicular transport.}
An undesirable property of structures formed in the pure VT limit is that 
they are quite sensitive to small variations in parameters and can even undergo an
 instability leading to runaway growth.  This instability is ameliorated by allowing
  fragments of cisternae (containing a \emph{finite fraction} $\alpha$ of cisternal mass)
 to bud off, move through the system and fuse with other compartments, as in the 
 cisternal progenitor model \cite{Pfeffer}. 
In fact, using flux balance arguments, we can show that the inclusion of sub-cisternal movement at any 
\emph{non-zero} rate $D$ is sufficient to eliminate runaway growth, and ensure that aggregates always 
attain a constant average mass even under substantial changes in the rates of transport processes (see SI).

 Moreover, if the flux of cisternal fragments is much lower than vesicular fluxes between cisternae, 
 then the well-separated $B$ and $C$ compartments typical of the pure VT model are preserved on an average (Fig. \ref{fig3a1}). 
 The movement of cisternal fragments does result in significant stochastic fluctuations ($\Delta m/m\sim \alpha$) 
 about the average mass profile over time scales proportional to $1/D$. These fluctuations, are however, 
 small as long as $\alpha$ is small. 
 Thus, the addition of a cisternal-progenitor-like process at a small rate to vesicular transport
appears to be a very reliable control mechanism, meeting all the desiderata of organelle morphogenesis.
In the rest of the paper, we refer to this strategy as the \emph{VT-dominated} case.

\paragraph*{Adding some vesicular exchange to pure cisternal progression.} 
Consider now a scenario where the bulk of transport is via cisternal movement, but with vesicular fission and exchange between
cisternae also occurring at a small rate.
As elaborated in \cite{Himani2013}, the system continues to support large aggregates as long as the the fission rate is small 
($\ll a/D$) and vesicular fluxes are independent of the mass of large parent aggregates. 
However, an interesting situation arises when  different species have very different fission rates, e.g., if  $w_{C}$ is large, while  $w_{A}$ and $w_{B}$ are small. In this case,
large aggregates form close to the cis end but disintegrate into smaller fragments near the trans end, giving rise to 
configurations (Fig. \ref{fig3b1}) that bear a striking resemblance to the usual picture of cisternal progression.
In the rest of the paper, we refer to this scenario as the \emph{CP-dominated} case.

Our analysis of mixed transport scenarios suggests that these are viable mechanisms for the formation and maintenance 
of differentiated cisternae, as long as one of the transport processes (vesicular or cisternal movement) is dominant, 
and the alternative process acts at a much smaller rate as a secondary track for traffic. 
If both processes occur with comparable rates, then large structures tend to dissipate, and the system  
 homogenizes.

\section*{Dynamical measurements can discriminate between transport models}
Below we discuss how experimental approaches that monitor dynamics of the Golgi in steady state or dynamics of recovery following perturbations from steady state, can be used to discriminate between 
various models. A dynamics-based approach was also explored in \cite{Patterson}, where the temporal pattern of fluorescence decay in iFRAP experiments was used to test several hypotheses about molecular trafficking in the Golgi (see  
\cite{Sens} for a quantitative analysis, also SI). Here, we propose several novel dynamical measurements that can provide a window into traffic processes.
\subsection*{Steady state dynamics of the Golgi} 
\begin{figure*}[h]
\vspace{-1cm}
\begin{minipage}{.3\linewidth}
\centering
\subfloat[]{\label{fig4a}
 \includegraphics[width=5.2cm,height=3.312cm]{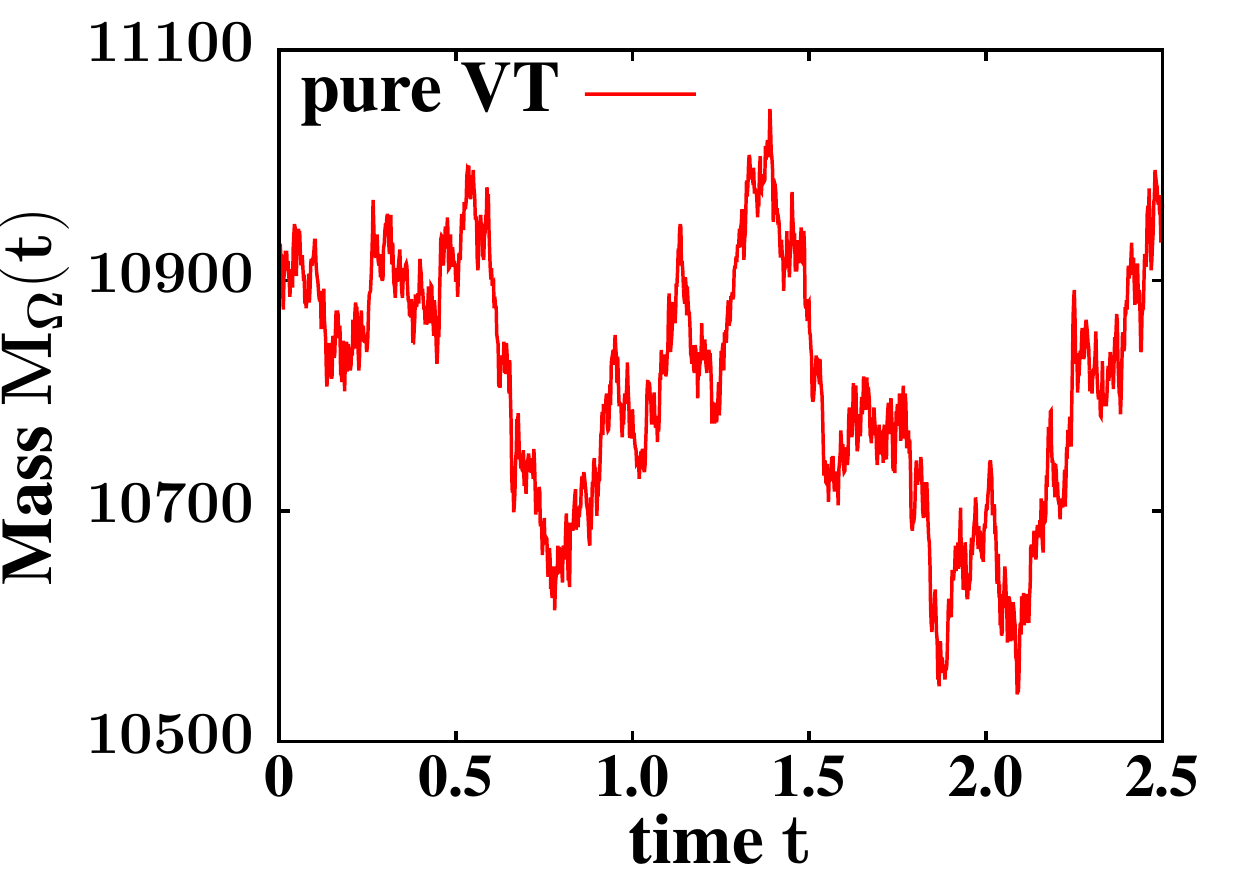}
  }\\
\subfloat[]{\label{fig4b}
  \includegraphics[width=5.2cm,height=3.312cm]{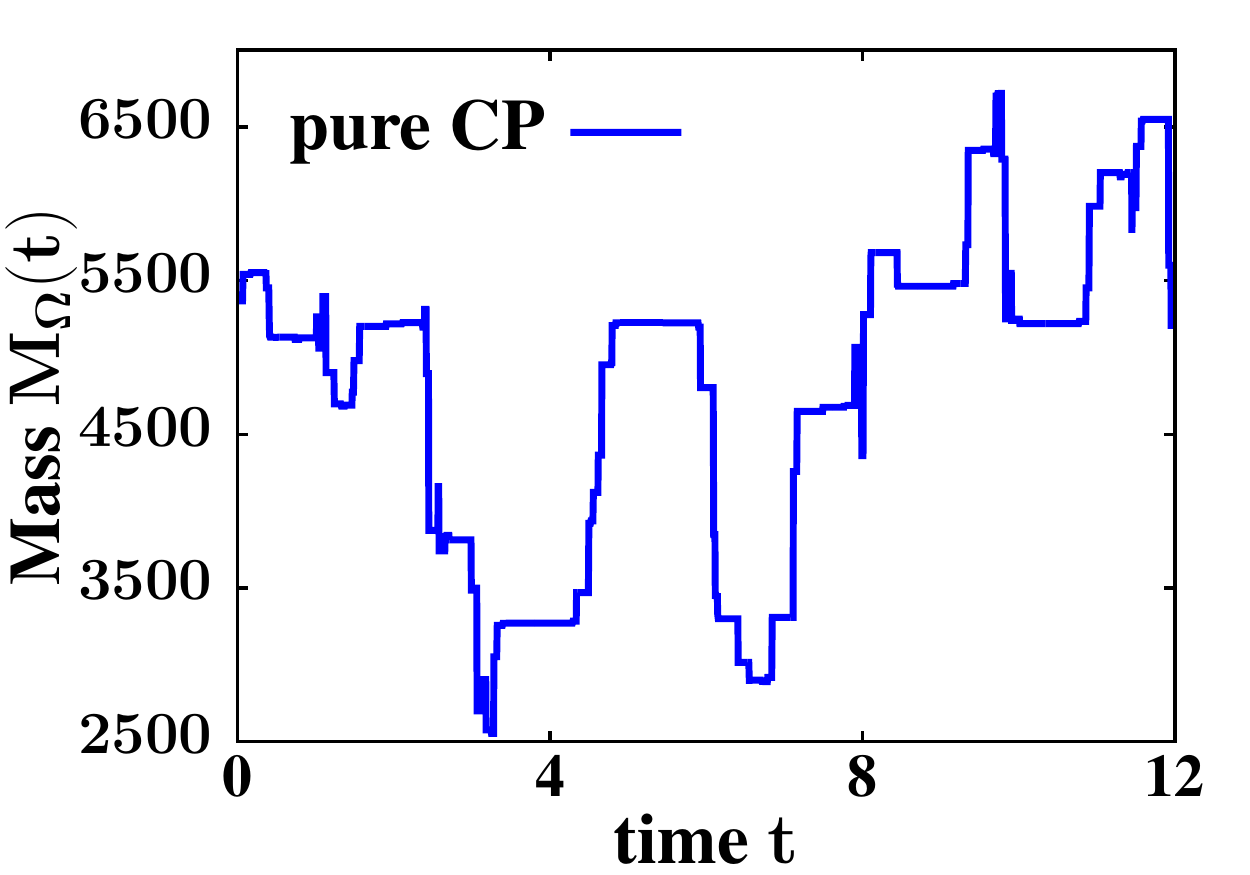}
  }
  \end{minipage}
\begin{minipage}{.3\linewidth}
\centering
\subfloat[]{\label{fig4c}
  \includegraphics[width=5.2cm,height=3.312cm]{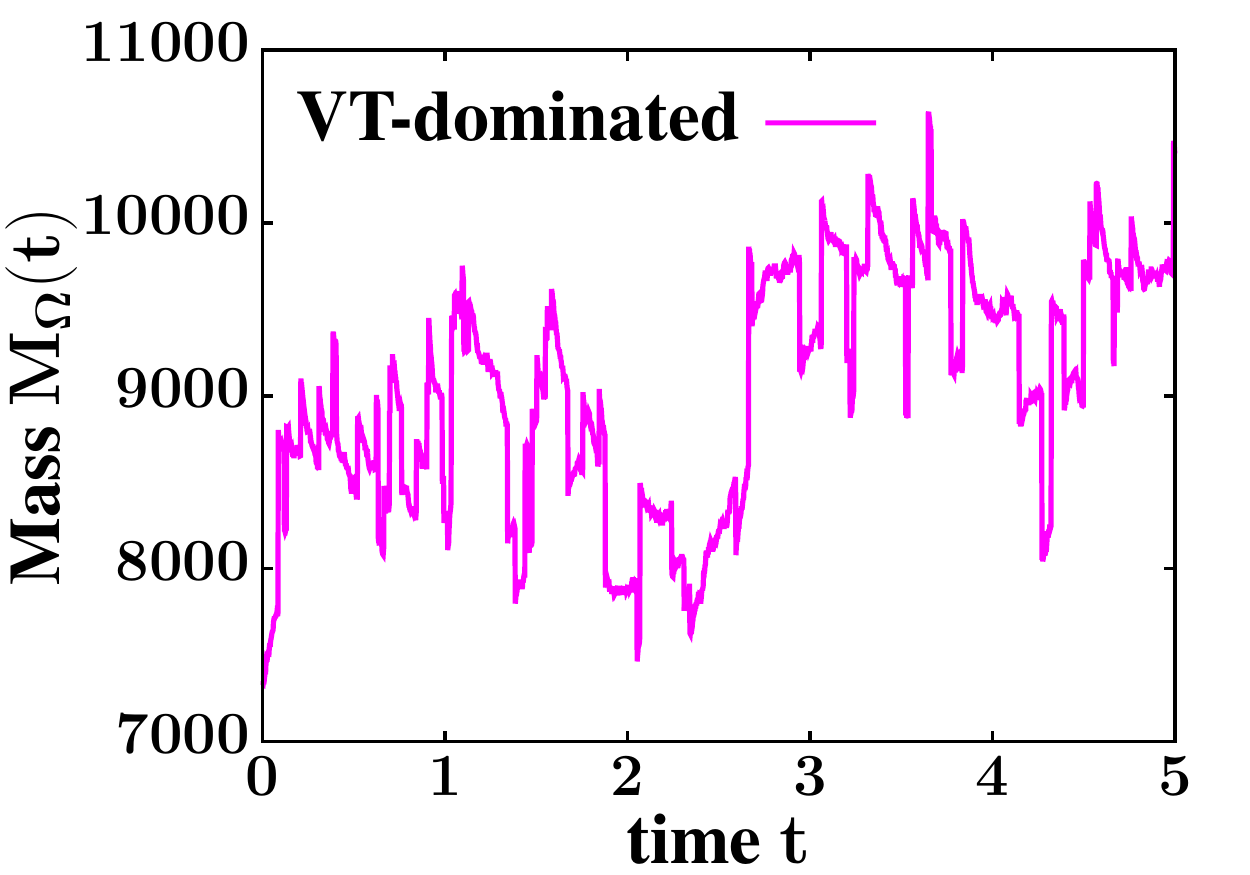}
  }\\
\subfloat[]{\label{fig4d}
  \includegraphics[width=5.2cm,height=3.312cm]{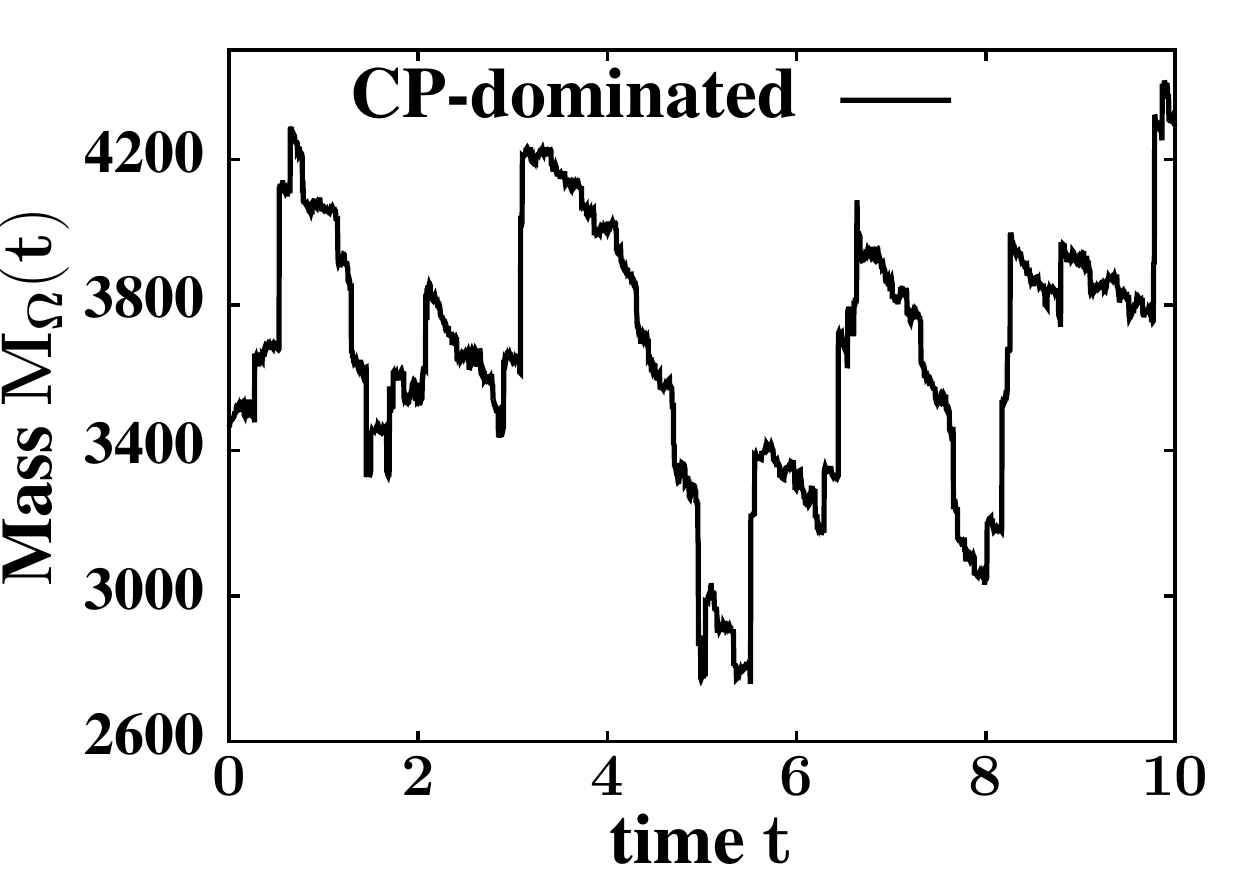}
  }
  \end{minipage} 
\begin{minipage}{.34\linewidth}  
\centering
  \subfloat[]{\label{fig4e}
 \includegraphics[width=5.6cm,height=3.9cm]{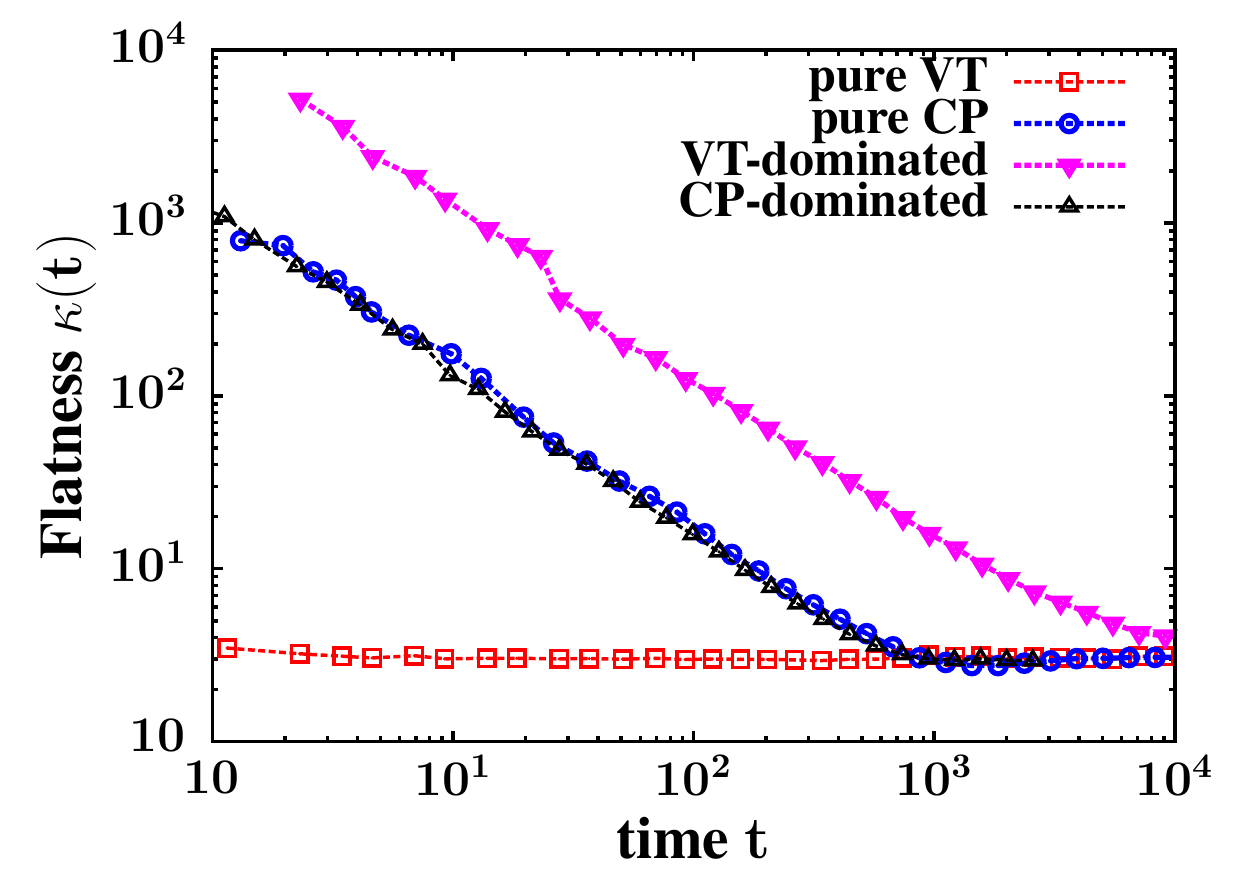}
  }\\
  \subfloat[]{\label{fig4f}
\includegraphics[width=5.6cm,height=3.9cm]{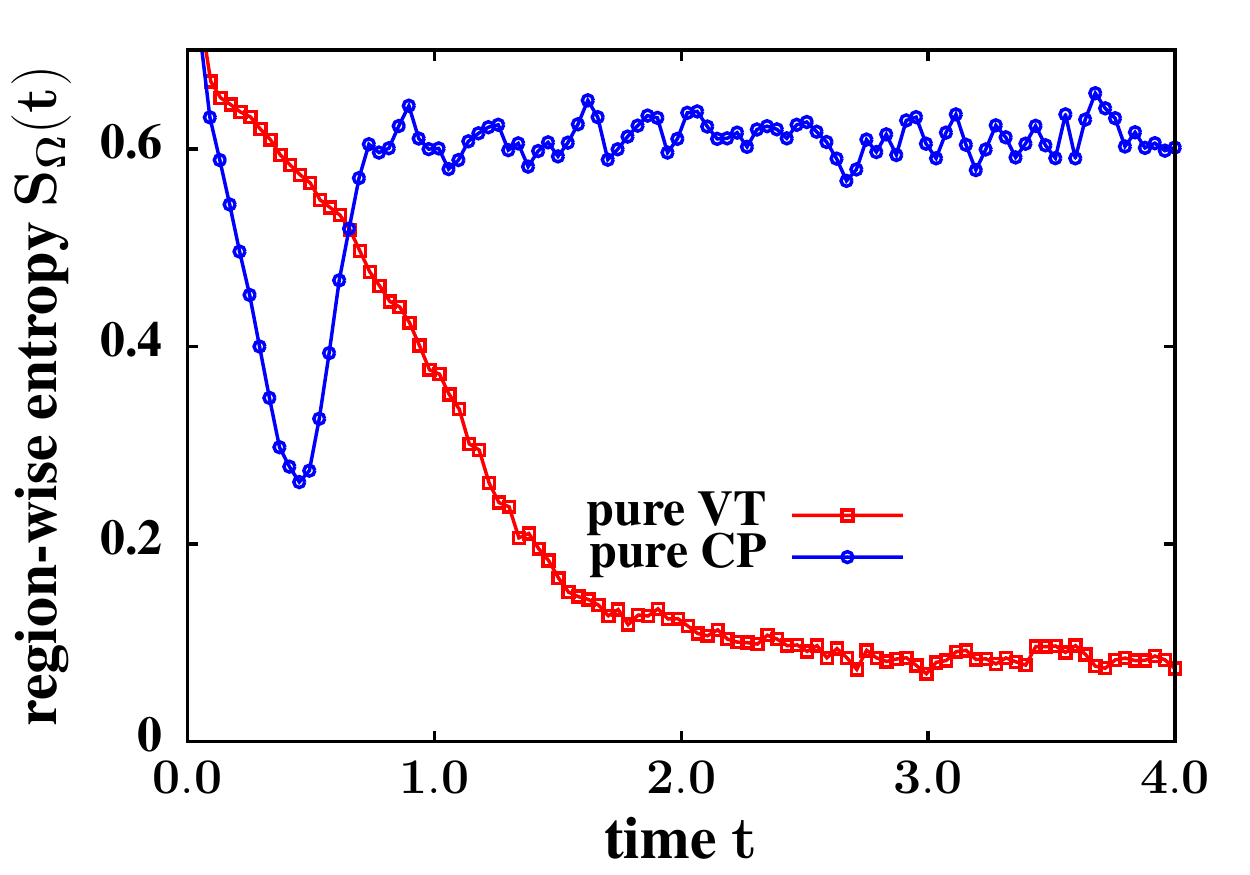}
}
\end{minipage}
  \caption{(a)-(d): Mass $M_{\Omega}(t)$ in the region ($0.25<x\leq 0.75 \in \Omega$) vs. time $t$ in steady state for (a) pure VT (b) pure CP 
 (c) VT-dominated model (with sub-cisternal movement) (d) CP-dominated model (with fission of C-type 
 vesicles). Scale of x-axis is $1\equiv 10^{5}$ t.u. for (a) and (c), and $1\equiv 10^{3}$ t.u. for (b) and (d).
Cisternal or sub-cisternal movement in cases (b)-(d) leads to sharp changes in $M(t)$.  
(e) Flatness $\kappa(t)$ vs. $t$ on a log-log plot. Sharp changes in 
$M(t)$ in cases (b)-(d) cause divergence of $\kappa(t)$ at small $t$. No such divergence with 
 pure VT. The time scale over which $\kappa(t)$ diverges is roughly the time interval between successive cisternal passage events, and is very larger for the VT-dominated case where sub-cisternal movement occurs rarely.  
(f) Dynamics during de novo biogenesis: Compositional entropy $S_{{\Omega}}(t)$ in the region  
($0.5<x\leq 0.75 \in {\Omega}$), averaged over $100$ realizations, decreases with $t$ for vesicle-dominated transport, and increases with $t$ for 
cisterna-dominated transport.
Scale of x-axis is $1\equiv 10^{5}$ t.u. in the pure VT and 
VT-dominated cases and $1\equiv 10^{4}$ t.u. in the pure CP and CP-dominated cases.
All simulation parameters same as in Fig. \ref{fig2m}.
}  
\label{fig4m}
\end{figure*}

Distinctive signatures of the
 underlying traffic mechanism can be observed in 
the time series and fluctuation statistics of the 
 ``mass'' $M_{\Omega}$ in (any) macroscopic 
 region $\Omega$ of the system.
Figures \ref{fig4a}-\ref{fig4d} depicts $M_{\Omega}(t)$ vs. $t$ 
for the central region $\Omega: 0.25<x\leq 0.75$ for the different models (see also SI movies (S1)-(S4)).
While the time series in the pure VT model has a smoothly varying, nearly Brownian character (Fig. \ref{fig4a}), 
 in the pure CP model, it is `intermittent' and displays sharp changes, corresponding to the entry and exit 
 of large aggregates from the region $\Omega$ (Fig.\,\ref{fig4b}). For the mixed transport, i.e., the 
 VT-dominated and the CP-dominated models,  the time series show stretches of smooth variation, 
 interspersed with an occasional sharp rise or fall (Figs.\,\ref{fig4d}).

 These observed differences can be quantified in terms  of  \emph{dynamical structure functions} 
 $S_{n}(t)=\langle [M_{\Omega}(t)-M_{\Omega}(0)]^{n}\rangle$ and their ratios, in particular the   
 `flatness'  \cite{Himani2013},
 \begin{equation}
  \kappa(t)=\langle [M_{\Omega}(t)-M_{\Omega}(0)]^{4}\rangle/\langle [M_{\Omega}(t)-M_{\Omega}(0)]^{2}\rangle^{2}\, ,
 \end{equation}
which is essentially a time-dependent kurtosis, the average $\langle \ldots \rangle$ being over several measurements of $M(t)$. If $M(t)$ shows variation on two very different scales,
 e.g.,  zero or small  changes from vesicular fluxes and very large changes from  cisternal or 
 sub-cisternal  passage, then $\kappa(t)$ is expected to show a divergence at small $t$, arising from the large changes \cite{Frisch, Himani2013}. 
We find that $\kappa(t)$
 diverges as $t\rightarrow 0$ for all the three cases - pure CP, CP-dominated, VT-dominated -
 in which there is appreciable cisternal or sub-cisternal movement, but remains `flat'  in the pure VT case 
(Fig.\,\ref{fig4e}). 
To further distinguish between the CP-dominated and VT-dominated scenarios, we monitor the \emph{timescale} over which flatness 
diverges-- this is just the typical time period between successive cisternal passage events, and is much larger for the VT-dominated limit, 
as compared to the CP-dominated limits (Fig. \ref{fig4e}). 
More generally, if the frequency of cisternal passage events (as inferred from flatness measurements) 
is consistent with the observed rates of cargo transit, then this points towards a cisterna-dominated mode of traffic, whereas
if the frequency is much less, then this would suggest that cisternal movement is only a secondary transport mechanism.



Statistics such as flatness can be extracted by monitoring the time series of fluorescence intensity of
 Golgi markers, and should be accessible in recent dynamical measurements at sub-Golgi resolution \cite{Tie_2016}.
 It may also be instructive to do this measurement separately for the cis, medial and trans regions of the Golgi, to test  
whether intensity fluctuations from cis and medial Golgi
have qualitatively different statistics compared to fluctuations  
 from the trans Golgi,  where transport is dominated by vesicles and fragments of the disintegrating cisterna. 


\subsection*{Dynamics of de novo biogenesis}
To study the dynamics of de novo biogenesis, we start with an initial condition in which the system is completely 
empty (corresponding to resorption of cargo molecules and Golgi-resident enzymes into the ER, for instance after 
Brefeldin treatment).  We then restart all the elementary processes that define the dynamics of the model (this corresponds to a washout of Brefeldin in regeneration experiments \cite{Langhans_denovo}), at the 
chosen rates, and monitor how the $A$, $B$, $C$ mass profiles recover in 
time $t$. Our simulations show that these profiles evolve quite differently, depending on
whether vesicular or cisternal movement is the dominant mode of transport (movies (S5)-(S6)). 

A clear signature of the transport process appears in the region-wise compositional `entropy', defined as:
\begin{equation}
S_{\Omega}=-\mathlarger{\mathlarger{\sum}}_{\nu=A,B,C,\ldots}\, \frac{M^{\nu}_{\Omega}}{M_{\Omega}}\log \left(\frac{M^{\nu}_{\Omega}}{M_{\Omega}}\right)
\label{eq3}
 \end{equation}
 where $M_{\Omega}^{\nu}$ is the mass of species $\nu$ in the region $\Omega$, and $M_{\Omega} = \mathlarger{\sum}_{\nu=A,B,C,\ldots}M_{\Omega}^{\nu}$.
The quantity $S_{\Omega}$ is a measure of the compositional heterogeneity 
of a compartment, being high for compartments with multiple species of particles, and low for compartments with 
predominantly one constituent type. The region $\Omega$ chosen for measurement must be of appropriate size-- neither so large that it spans multiple compartments,
nor so small that the $M_{\Omega}(t)$ signal is dominated by noise.
 When cisternal movement is the predominant mode of transport, the region-wise entropy $S_{\Omega}(t)$ 
  increases with $t$ as the system regenerates, in contrast to the pure VT and VT-dominated limits, where
$S_{\Omega}(t)$ is a decreasing function of $t$, except for an initial transient rise (Fig.\,\ref{fig4f}).

In the CP limit, rising entropy in the cis compartments is a direct consequence of maturation-- the early aggregate is purely of type A, but changes from A to AB to ABC to BC over time, thus increasing the entropy in this region.
Near the trans end, maturation has the opposite effect, shifting the composition of aggregates from BC towards pure C and reducing the entropy. But this is countered by  the entry of mixed BC compartments
and exit of concentrated C compartments, resulting in a net increase in compositional entropy even in the trans compartments, on average.
Thus, $S_{\Omega}(t)$ rises during de novo formation with CP due to the generation and movement of \emph{mixed} compartments.
By contrast, in the VT limit, new $C$ compartments are not generated by the maturation of $B$ compartments via mixed $BC$ intermediates, but instead arise due 
 to the preferential localization of $B$ and $C$ particles in different regions.
The long period of decreasing 
entropy thus corresponds to growth of $B$ and $C$ particles in different regions and the increase in the `purity' of the 
two compartments. 

In the SI, we also describe the dynamics of reconstitution (from an initial state where all molecules and markers
are distributed in the Golgi region in a uniform, unpolarized manner) and discuss how this differs from de novo formation.


\subsection*{Perturbation and response}
We now consider how the steady state structures in the various models respond to perturbations such as 
(i) variation in influx and (ii) exit block at the trans end.

\begin{figure}[h]
\centering
\includegraphics[width=0.45\textwidth]{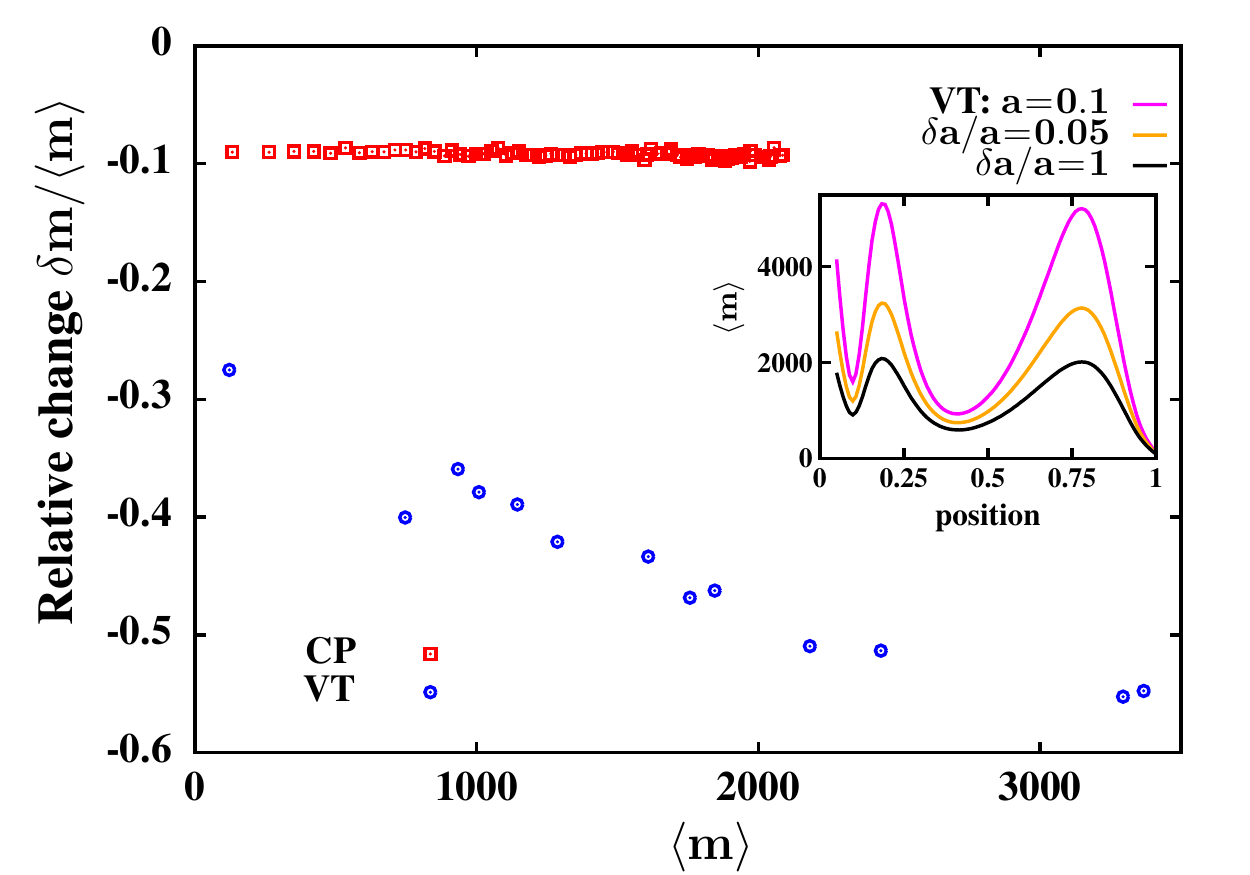}
\caption{Relative change $\delta m/\langle m\rangle$ (at a location with average local mass $\langle m\rangle$) when influx $a$ falls by 10\%. In the CP limit, $\delta m/\langle m\rangle$ is independent of $\langle m\rangle$, while in the VT limit, $|\delta m|/\langle m\rangle$ increases with $\langle m\rangle$: larger
masses show a stronger proportionate change than smaller masses. Inset: In the VT limit, mass profiles  become nearly flat when $a$ is reduced or very sharply peaked when $a$ increases. 
}
\label{vary_a}
\end{figure}
(i) Change in influx: 
How does the average 
mass $\langle m\rangle$ at any location within the Golgi respond when the ER to Golgi influx shifts from $a$ to $a\pm\delta a$?
Fig.\ref{vary_a} shows the relative change $\delta m/\langle m\rangle$ as a function of  $\langle m\rangle$ for the 
pure VT and pure CP models, for $\delta a/a=-0.1$. In the CP model, $\delta m/\langle m\rangle$ is independent of 
$\langle m\rangle$, indicating that all cisternae, whether large or small, show the same proportionate reduction in volume on an average.
By contrast, in the VT model,  $|\delta m|/\langle m\rangle$ increases with $\langle m\rangle$.
Thus, the relative change is maximum where local mass is highest and minimum for regions with low mass,
resulting in a new mass profile that is less peaked (inset, Fig.\ref{vary_a}). 
The strong, shape-altering response 
of the mass profile in the VT limit is a direct consequence of the flux-kernel $f[m]$ saturating at 
large $m$. Large aggregates, which operate in this saturation regime, must undergo a substantial change in size in order to 
produce the small changes in vesicular outflux required to balance the altered influx.  
Note that while cisternal sizes respond to altered influx in both VT and CP limits, it may be difficult to distinguish a systematic change from large stochastic fluctuations in the pure CP limit without sufficient  
statistical averaging (see also movies (S9)-(S12)). 
\paragraph*{}
(ii) Exit block: Golgi transport is highly temperature-sensitive, with budding and exit from the trans Golgi suppressed at low temperatures ($\sim 20^{\circ} C$), resulting in lateral expansion of the trans compartments
\cite{Ladinsky}. We simulate the $20^{\circ} C$ block by stopping the exit of particles from the right boundary, and find that particles pile up indefinitely at the exit site in both the CP and VT limits (see movies (S13)-(S14)). 
This uncontrolled pile-up is unrealistic for the Golgi, but raises the interesting question of how the size of trans compartments is regulated when exit is blocked-- whether there is an alternative pathway for the degradation of the trans Golgi
 or some feedback mechanism that eventually suppresses influx from the ER.

\section*{Discussion}
\label{sec:sec5.7}

In this paper, we have defined and studied a general but minimal stochastic model of a system of compartments
within a trafficking pathway. While we have chosen to focus on the Golgi apparatus, the model is equally relevant to
the endosomal system. Our model
is broad enough to survey several known local dynamical rules of transport and chemical transformation, and
encompasses vesicular transport, cisternal progression, cisternal progenitor and its variants.
We identified conditions
under which this nonequilibrium open system self-organizes into Golgi-like structures, i.e., into multiple, spatially resolvable 
aggregates with a position-dependent chemical composition. We emphasize that 
the models studied here perforce have a many-body character arising from interactions between the constituent particles, this implies that {\it local} changes in parameter values can have {\it nonlocal} phenotypic consequences.


In the context of vesicular transport, we have emphasized (as has \cite{Sens}) the importance of the MM-form of the flux kernel 
$f[m]$ for fission and chemical conversion processes, which shows saturation at large $m$.
Experimentally probing saturation effects in  vesicular fluxes as compartment size is varied, can
be an interesting direction for future work, and may also have broad
implications for formation and maintenance of sub-cellular structures.

Significantly, our model generates chemically polarized compartments
even in the absence of directed and specific fusion, an important ingredient of previous models \cite{Sens, Heinrich}. 
In these models, directed fusion essentially results in autocatalytic accumulation of a particular chemical species in 
a particular compartment. This appears to be necessary for formation of non-identical compartments in a situation
where the dynamics is completely symmetric \cite{Heinrich}. However, our model suggests that in the scenario where there is an inherent 
asymmetry or directionality, as in the case of the Golgi apparatus, specific fusion may be less central to the 
maintenance of compartments with distinct chemical identities.

A limitation of this study is that it does not explicitly take into account the role of enzymes in the processing 
of cargo molecules in the Golgi, and the complex dynamics of the enzymes themselves. Thus, extending the present model to
include enzymes as separate entities would lead to a more realistic model, the parameters of which could also be connected 
to experimentally relevant quantities. 
However, the increase in the number of parameters also makes it difficult to analyze such an extended model and extract the relevant 
regions of parameter space. An abstract model of the sort studied in this paper, is more suited to gain
insight into the qualitative effects at play in the system. 
At a philosophical level, our work touches upon the question of whether and how the Golgi can form de novo 
(for example, after treatment with reagents that cause Golgi cisternae to disintegrate).
We show that a system with ``Golgi-like traffic processes'' has the ability to self-organize into morphological and chemically distinct
structures that share many 
qualitative features of  Golgi cisternae. 
This self-organization happens even in the limit corresponding to vesicular transport, in the absence of any template.
In essence, the molecular machinery associated with the stated dynamical rules of the minimal models are the only ones necessary for
Golgi inheritance.
 Indeed, this and many of our mathematically derived conclusions find resonance
in ideas articulated before, especially in \cite{glickmalhotra1998}.
An important conclusion of our analysis is that the transport mechanism can qualitatively impact 
structural features of the cisternae such as inter-cisternal tubulations or the degree of localization of Golgi-resident 
enzymes. Such correlations between transport and morphology could be explored by doing a comparative study across
different cells.
Finally, we propose that fluctuation statistics in fluorescence experiments could be a powerful probe of Golgi dynamics,
with sporadic and large fluctuations, in particular, revealing cisternal passage events.
\paragraph*{Acknowledgments---}
HS thanks NCBS for hospitality. We thank Vivek Malhotra and Mukund Thattai for critical discussions and suggestions.

\newpage

\begin{center}
\large{Supplementary Online Material for}
\vskip 0.2in
\Large{\bf Nonequilibrium mechanisms underlying de novo biogenesis of Golgi cisternae}
\vskip 0.2in
\Large{Himani Sachdeva$^{1,2}$, Mustansir Barma$^{1,3}$ and Madan Rao$^{4,5}$\\
\vskip 0.1in
\normalsize{$^{1}$}Department of Theoretical Physics, Tata Institute of Fundamental Research, Homi Bhabha Road,
Mumbai 400005, India\\
\normalsize{$^{2}$}Institute of Science and Technology, Am Campus 1, Klosterneuburg A-3400, Austria\\
\normalsize{$^{3}$}TIFR Centre for Interdisciplinary Sciences, 21 Brundavan Colony, Narsingi, Hyderabad 500075, India\\
\normalsize{$^{4}$Raman Research Institute, C.V. Raman Avenue, Bangalore 560080, India}\\
\normalsize{$^{5}$Simons Centre for the Study of Living Machines, National Centre for Biological Sciences (TIFR), Bellary Road, Bangalore 560065, India}\\
\vskip 0.1in
}

\end{center}
\tableofcontents

\section{Model}
\subsection{Detailed Description}
The model is a 1D lattice model  with three species, `A', `B' and `C' of particles (see Fig.1 of main paper for a schematic).
It incorporates the main processes involved in Golgi dynamics as stochastic moves (listed below) that occur with a constant rate per unit time.   
\begin{enumerate}
\renewcommand{\theenumi}{\roman{enumi}}
\item \emph{Influx} of single particles of type A at the leftmost site (ER) with rate $a$.
\item \emph{Chemical conversion} of an A particle to a B particle at site $i$  with rate $uf[m^{A}_{i}]$
that depends on the ``mass'' (number of particles)  $m^{A}_{i}$ of A through the flux-kernel $f[m^{A}_{i}]$.
Similarly, a $B\rightarrow C$ conversion can occur at site $i$ with rate $vf[m^{B}_{i}]$, and so on
\footnote{In principle, the reactions $B\rightarrow A$ and $C\rightarrow B$ can also be included, but to limit the number
of parameters, we do not allow for these. The basic features of the model are not altered in the presence of these 
reverse reactions, as long as there is a net forward rate of reaction.}.
\item \emph{Fission, movement and fusion of single particles} with species-dependent rates: An $A$ particle  
fissions from site $i$, moves in the anterograde (or retrograde) direction with rate $w_{A}f\left[m^{A}_{i}\right]$ 
(or $w^{\prime}_{A}f\left[m^{A}_{i}\right]$), and fuses with the mass on the neighboring site. Similarly a B (or C) particle fissions and moves forward with rates 
$w_{B}f\left[m^{B}_{i}\right]$ (or $w_{C}f\left[m^{C}_{i}\right]$) and backward with rates 
$w^{\prime}_{B}f\left[m^{B}_{i}\right]$ (or $w^{\prime}_{C}f\left[m^{C}_{i}\right]$).
\item  \emph{Breakage  of a finite fraction ($\alpha$) of an aggregate and movement} in the anterograde direction with rate $D$ 
($\alpha=1$ corresponds to cisternal progression, and $\alpha<1$ to sub-cisternal movement.).
\item \emph{Exit of full aggregates or single particles from boundaries}, with the same rates as those for movement 
in the bulk \footnote{If exit is allowed to occur only from the trans end, then we need to consider 
$\mathcal{O}\!(1/L)$ injection rates.}.
\end{enumerate}
\paragraph*{}

The vectorial nature of transport is modeled through the asymmetry in the anterograde/retrograde particle movement rates, and is parametrized as: 
$w_{A}\rightarrow \gamma_{A} w_{A}$, $w^{\prime}_{A}\rightarrow \left(1-\gamma_{A}\right) w_{A}$, 
$w_{B}\rightarrow \gamma_{B} w_{B}$, $w^{\prime}_{B}\rightarrow \left(1-\gamma_{B}\right) w_{B}$ and 
$w_{C} \rightarrow \gamma_{C} w_{C}$, $w^{\prime}_{C}\rightarrow \left(1-\gamma_{C}\right) w_{C}$ (see Fig. 1 of main paper).
The asymmetry factors $\gamma_{A}$, $\gamma_{B}$, $\gamma_{C}$ for the three species can be different,
representing differential degrees of 
recycling of $A, B, C, \ldots$  particles to the ER (a generalization of  \cite{Himani2011}).

The parameter space of
this model is quite large, encompassing the injection rate $a$, the interconversion rates $u, v, \ldots$
the fission rates $w_{A}, w_{B}, w_{C}, \ldots$ and the corresponding asymmetry factors $\gamma_{A}, \gamma_{B}, \gamma_{C}, 
\ldots$, the aggregate movement rate $D$, the breakage fraction $\alpha$, and finally the form of the function $f$ itself.

\subsection{Rationalizing the model}
Below we comment in some detail on various aspects of the model and also highlight its strengths and weaknesses as a model of 
Golgi biogenesis.
\begin{enumerate} 
\item Representing the three-dimensional Golgi by a one-dimensional model:\\
Our model is a spatial model that explicitly incorporates distance from the cis end as a relevant variable. 
In the interest of analytical tractability, we include only one spatial dimension corresponding to the cis to trans direction, which
 is the main direction of molecular traffic, and also the axis along which the Golgi exhibits biochemical polarity. 
The 1D model can thus be considered an effective model obtained by integrating over the two directions perpendicular to this cis to trans
direction. However, a 1D model of this sort cannot address questions related to the shape  of 
individual cisternae or if trafficking pathways are branched.

\item A,B,C particles represent molecules in different stages of processing in the Golgi: \\
In constructing this model, we imagine an $A$ particle to be the equivalent of the unprocessed protein molecules arriving 
at the cis-Golgi from the ER, and particles of type $B$, $C$, $D$ as representing proteins in different (successive) 
stages of processing. Likewise, an aggregate which is primarily of type $B$ is analogous to a cisterna with a large fraction of 
semi-processed molecules in an early stage of processing.
Thus, in our model, the processing stages `A', `B', `C' of the primary constituent molecules of a cisterna 
define the chemical identity of the cisterna. 

\item Modeling enzyme-mediated cargo modifications as simple Poisson processes:\\
The $A\rightarrow B$ and $B\rightarrow C$ conversions which we have treated as Poisson rate processes in the model are
catalyzed by enzymes in the real system. A more realistic model of the Golgi could include both $A$, $B$, $C$ 
`cargo species' and $E_{A}$, $E_{B}$ `enzyme species', with specific enzymes having an affinity for specific cargo types. 
This kind of a detailed enzyme plus cargo model would allow us to model feedbacks wherein the distribution of $B$ particles influences the distribution of the
  corresponding $E_{B}$ enzymes, and is in turn influenced by $E_{B}$ molecules which activate the $B\rightarrow C$ conversion.
Thus, in general, we expect self-organization in this extended model to be more complex.
Nevertheless, the effective rate-based model studied in this paper incorporates the sequential interconversion process using 
just a few parameters, and has the advantage that it is more amenable to analysis than a complex model.
Moreover, as a \emph{spatial} model that incorporates both chemical and transport processes in the Golgi, 
it provides a valuable framework for constructing a more detailed model with enzymes.
\item Model parameters are composites of many biophysical rates:\\ The various particle exchange and interconversion rates in our model are effective or 
composite rates. For example, the particle movement rate combines both the rate at which a vesicle buds from an aggregate, as well as the rate at which it
fuses with the next aggregate.
Similarly, the $B\rightarrow C$ modification rate is determined jointly by the rates of attachment and detachment of 
enzyme molecules with B molecules, the concentration of the enzyme molecules, the reaction rate between the enzyme and the B molecule etc. . Because of the composite nature of the rates, it is not straightforward to find the 
corresponding experimentally measurable parameter. However, 
a model of this sort with a relatively small number of composite parameters 
allows for an economical description of the system, which makes it easier to identify and
elucidate the qualitative effects at play in the system. 
\end{enumerate}

\subsection{Dynamical equations for mass}
The following dynamical equations describe the time evolution of the average mass (number of particles) of each species at 
each site $i$ in the lattice, in accordance with the elementary moves allowed in the model:
\begin{subequations}
 \begin{equation}
\frac{\partial \langle m^{A}_{i}(t)\rangle}{\partial t} = w_{A}\left\{\gamma_{A}\langle f[m^{A}_{i-1}] \rangle +(1-\gamma_{A})\langle 
f[m^{A}_{i+1}] \rangle-\langle f[m^{A}_{i}] \rangle\right\}+ \alpha D\left\{\langle m^{A}_{i-1} \rangle -\langle m^{A}_{i} \rangle\right\} - u\langle f[m^{A}_{i}] \rangle
 \label{eqn:eq1a}
\end{equation}
 \begin{equation}
\frac{\partial \langle m^{B}_{i}(t)\rangle}{\partial t} =  w_{B}\left\{\gamma_{B}\langle f[m^{B}_{i-1}] \rangle +(1-\gamma_{B})\langle 
f[m^{B}_{i+1}] \rangle-\langle f[m^{B}_{i}] \rangle\right\}+\alpha D\left\{\langle m^{B}_{i-1} \rangle -\langle m^{B}_{i} \rangle\right\}
 + u\langle f[m^{A}_{i}] \rangle-k\langle f[m^{B}_{i}] \rangle
\label{eqn:eq1b}
\end{equation}
 \begin{equation}
\frac{\partial \langle m^{C}_{i}(t)\rangle}{\partial t} = w_{C}\left\{\gamma_{C}\langle f[m^{C}_{i-1}] \rangle +(1-\gamma_{C})\langle 
f[m^{C}_{i+1}] \rangle-\langle f[m^{C}_{i}] \rangle\right\}+\alpha D\left\{\langle m^{C}_{i-1} \rangle -\langle m^{C}_{i} \rangle\right\} 
+ k\langle f[m^{B}_{i}] \rangle
\label{eqn:eq1c}
\end{equation}
\begin{equation}
w_{A}\gamma_{A}\langle f[m^{A}_{0}] \rangle=a; \quad \langle f[m^{B}_{0}] \rangle=\langle f[m^{C}_{0}] \rangle=0; \quad 
\langle f[m^{A}_{L+1}]\rangle= \langle f[m^{B}_{L+1}]\rangle=\langle f[m^{C}_{L+1}] \rangle=0 
\end{equation}
\label{eqn:eq1}
\end{subequations}
where $\langle \ldots\rangle$ indicates averaging over ensembles.
The mass of each species at each site changes due to 
single particle exchange and aggregate movement between neighboring sites, as well as interconversion at any given site. If the system attains 
steady state, then the time derivatives in eqs. \eqref{eqn:eq1a}-\eqref{eqn:eq1c} can be set to zero, so that the equations take the form of 
flux-balance conditions for each site.

\section{Analysis of the pure Vesicular Transport (VT) case}
\subsection{Insights from the analytical solution}
\label{sec:sec2.1}
\begin{figure}[h]
\subfloat[] {
\centering
\includegraphics[width=6.6cm,height=4.53cm]{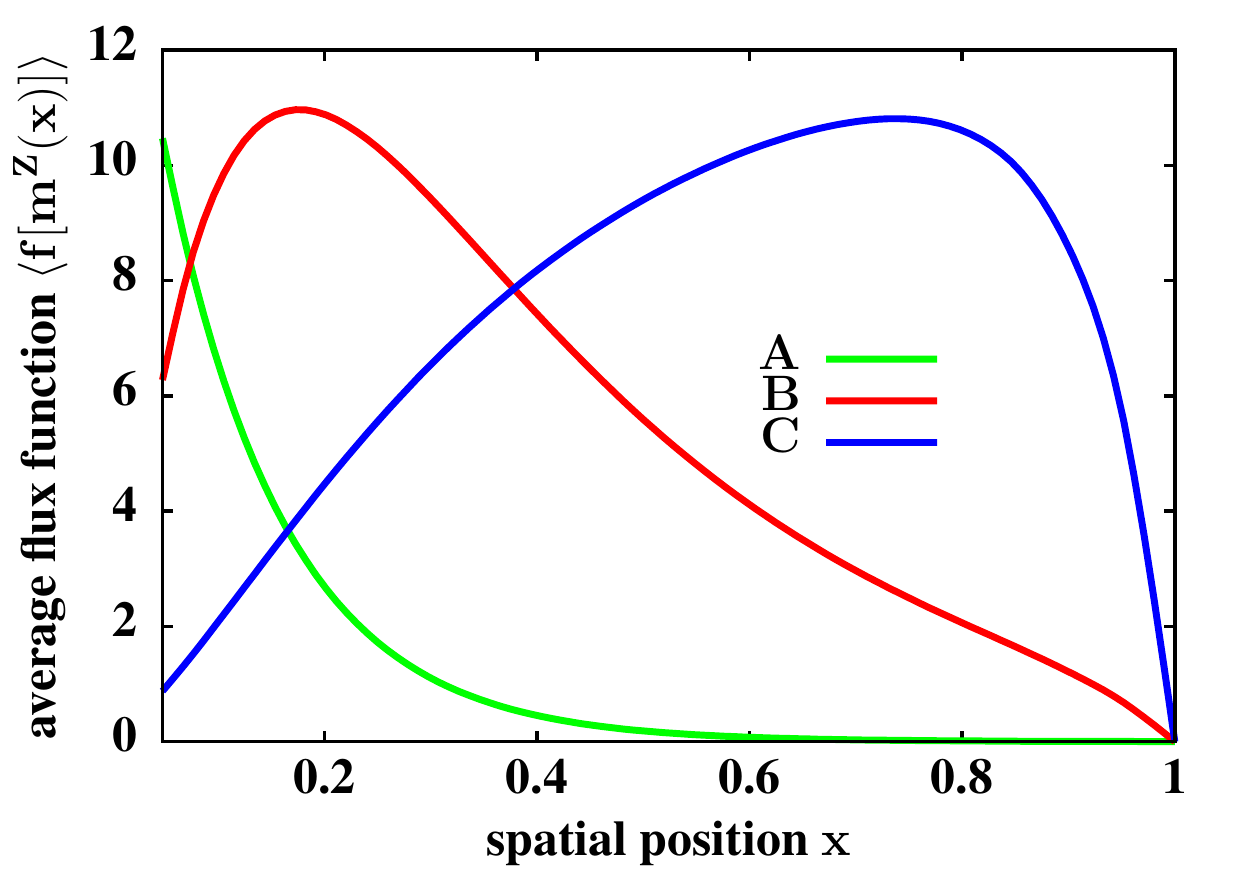}
\label{fig1a}}
\subfloat[]{
\includegraphics[width=6.6cm,height=4.53cm]{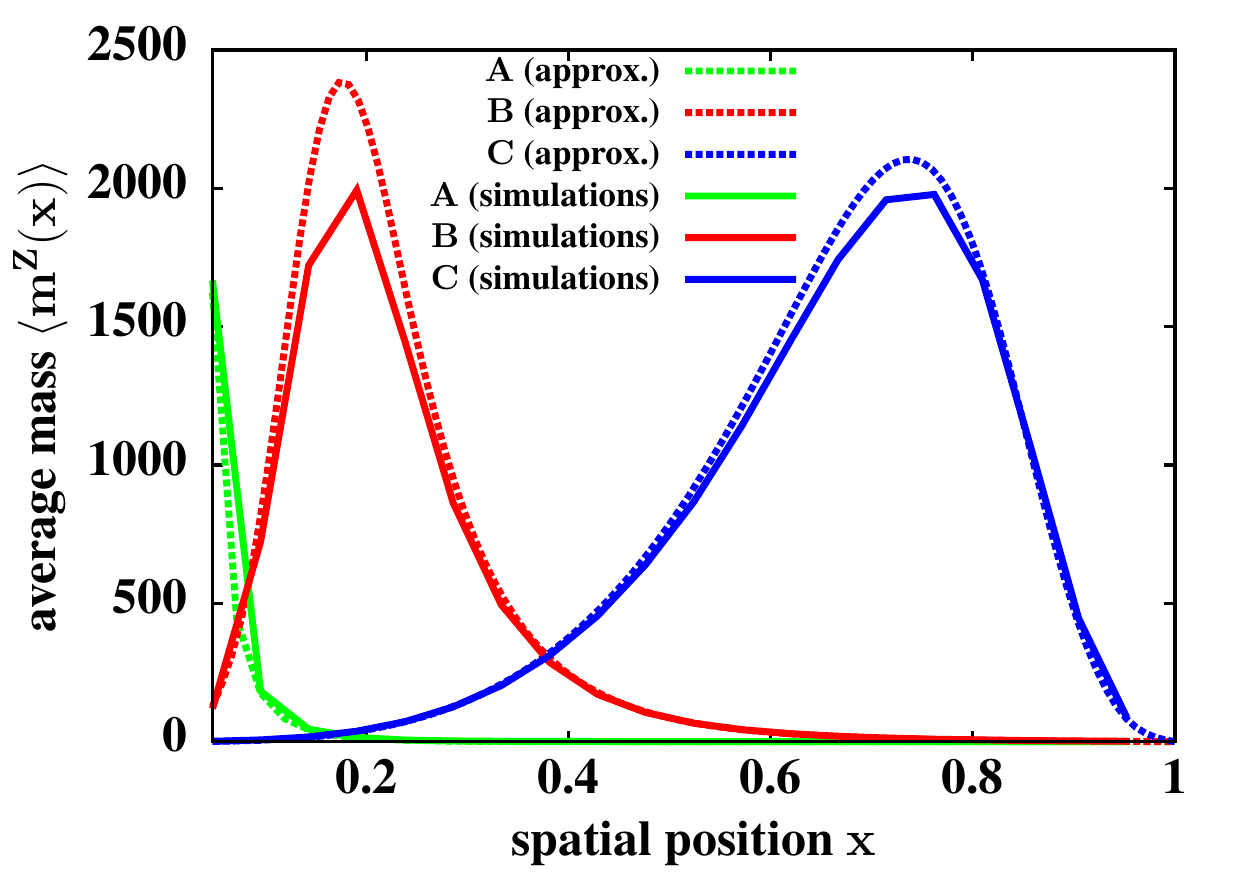}
\label{fig1b}}
\caption{(a) Equation \eqref{eqn:eq1} can be solved numerically in the continuum limit to obtain $\langle f[m^{A}(x)]\rangle$,
$\langle f[m^{B}(x)]\rangle$ and $\langle f[m^{C}(x)]\rangle$ vs. $x$ (here,   
for $a=1$, $D=0$, $w_{A}=0.125$, $w_{B}=0.04375$, $w_{C}=0.05462$, $\gamma_{A}=0.5$, $\gamma_{B}=0.6$, 
 $\gamma_{C}=0.66$, $u=0.01125$, $v=0.00194$, $m_{sat}=200$, $K_{sat}=14.14$).
(b) The approximate $\langle m^{A}(x)\rangle$, $\langle m^{B}(x)\rangle$, $\langle m^{C}(x)\rangle$
 profiles (dashed lines) derived using eq. \eqref{eqn:eq2b}  are reasonably close to the actual mass profiles (solid lines) from simulations. 
 }
\label{fig1}
\end{figure}

\paragraph*{}
In the pure VT limit (with $D=0$), eq. \eqref{eqn:eq1} can be solved in the steady state, 
by setting time derivatives to zero and taking a continuum limit $i/L\rightarrow x$ in space, which transforms the equations
into a set of coupled, second-order ODEs. Solving these ODEs yields $\langle f[m^{A}(x)]\rangle$, 
$\langle f[m^{B}(x)]\rangle$ and $\langle f[m^{C}(x)]\rangle$ as a function of $x$. 
For the three-species model, the solutions are quite involved, and it is more convenient to solve 
 eq.  \eqref{eqn:eq1} numerically. Figure \ref{fig1}\subref{fig1a} shows numerically obtained solutions 
 for a particular choice of parameters. The mass profiles $\langle m^{A}(x)\rangle$, $\langle m^{B}(x)\rangle$ and 
 $\langle m^{C}(x)\rangle$ can be approximately derived from this solution by assuming that the mass $m$ at any instant is close to its average value $\langle m\rangle$, so that 
$\langle f[m]\rangle\sim f[\langle m\rangle]$.
For instance, if $f[m^{Z}]$ is of the Michelis-Menten (MM) form [eq. \eqref{eqn:eq2a}] (where $Z$ can denote any of $A$, $B$, $C$), then the mass profiles can be approximated  using eq. \eqref{eqn:eq2b} [see also fig. \ref{fig1}]. 
\begin{subequations}
\begin{equation}
 f[m^{Z}]=\frac{K_{sat}\sqrt{m^{Z}}}{\sqrt{m^{Z}}+\sqrt{m_{sat}}} 
\label{eqn:eq2a}
\end{equation}
 \begin{equation}
 \langle m^{Z}\rangle\sim \left( \frac{\sqrt{m_{sat}}\langle f[m^{Z}]\rangle}{K_{sat}-\langle f[m^{Z}]\rangle}\right)^{2}
 \label{eqn:eq2b}
 \end{equation}
\label{eqn:eq2}
\end{subequations}

 \paragraph*{}
The key characteristics of the mass profiles can be illustrated using a two-species version of the model with  
`A' and `B' particles that undergo $A\rightarrow B$ conversion, and move with the same directional bias 
$\gamma_{A}=\gamma_{B}=\gamma$. In this case, the analytical steady state solution for $\langle f[m^{A}(x)]\rangle$ and 
$\langle f[m^{B}(x)]\rangle$ is relatively simple \cite{Himani2011}:
\begin{subequations}
 \begin{equation}
\langle f[m^{A}(x)]\rangle  =\frac{a}{w_{A}\gamma}
\exp(\tilde{\xi} x) \left(\frac{\sinh\left(\sqrt{\tilde{\eta} + \tilde{\xi}^{2}}(1-x)\right)}{\sinh\left(\sqrt{\tilde{\eta} + \tilde{\xi}^{2}}\right)}\right), \qquad  \tilde{\eta} = \frac{2uL^{2}}{w_{A}}, \quad \tilde{\xi} = (2\gamma - 1)L
 \end{equation}
\begin{equation}
\langle f[m^{B}(x)]\rangle =\frac{a}{w_{B}\gamma}\Vast[\frac{1-\exp(-2\tilde{\xi}(1-x)) }{1-\exp(-2\tilde{\xi})}
  -\exp(\tilde{\xi} x) \left(\frac{\sinh\left(\sqrt{\tilde{\eta} + \tilde{\xi}^{2}}(1-x)\right)}{\sinh\left(\sqrt{\tilde{\eta} + \tilde{\xi}^{2}}\right)}\right)\Vast]
\end{equation}
\label{eqn:eq3}
\end{subequations}
\begin{figure}
\begin{center}
\includegraphics[width=0.4\textwidth]{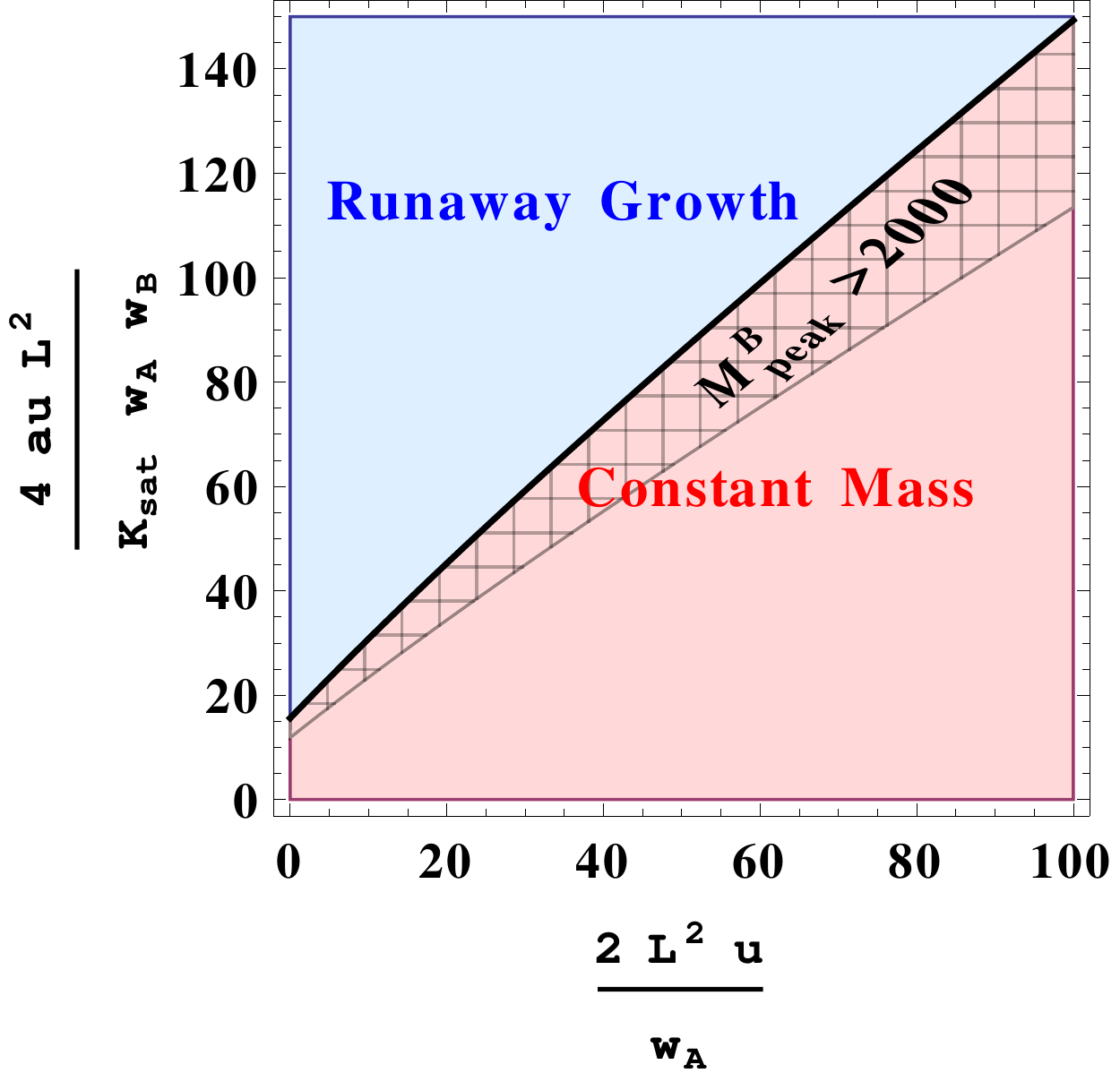}
\end{center}
\caption{Phase diagram for the simplified two-species model with $\gamma_{A}=\gamma_{B}=1/2$, $2a/w_{A}<1$ and $f[m]$ of the MM type. 
Parameter combinations in the blue region lead to runaway growth of B particles; parameters in red region correspond to steady state, and the bold line is the boundary between the two phases.
For parameters in the shaded red region, $\langle m^{B}\rangle>2000$ at the peak of the B profile (as deduced from the approximation in eq. \eqref{eqn:eq2b}).}
\label{fig2}
\end{figure}
 Below, we discuss some salient features of this solution:
\begin{enumerate}
 \item \emph{Stationary state vs. runaway growth}: Consider a flux kernel $f[m]$ of the sort in eq. \eqref{eqn:eq2a},
 which saturates to a constant value  $K_{sat}$ at large $m$. The average  $\langle f[m^{A}(x)]\rangle$ and $\langle
 f[m^{B}(x)]\rangle$ functions, as  given by eq. \eqref{eqn:eq3}, must be less than this maximal value $K_{sat}$ everywhere, for eq. \eqref{eqn:eq3} to be the correct steady state solution.
For some parameters, however, unphysical solutions arise with 
  $\langle f[m^{A}(x)]\rangle$ and/or $\langle f[m^{B}(x)]\rangle$ exceeding $K_{sat}$. These parameters are precisely those 
  for which the assumption of steady state ($d\langle m\rangle/dt=0$) breaks down; instead, the mass undergoes runaway growth: $d\langle m\rangle/dt>0$ at all times (for details, see \cite{Himani2011}). 
 Figure \ref{fig2} depicts a phase diagram for the 2-species model with $\gamma=1/2$, where parameter combinations in the blue and red regions lead to  
runaway growth of B particles and steady mass of B particles respectively.
  Physically, runaway growth arises 
   because the outgoing vesicular fluxes at a site saturates if the mass at the site is much higher than $m_{sat}$. 
 Thus, at sufficiently high rates of influx, the outgoing flux from a location may fail to balance the incoming flux, 
 leading to runaway growth.
   A more detailed analysis shows that typically, runaway growth occurs only in specific regions of the system, even as
   the rest of the system attains steady state \cite{Himani2011}. 
 \item  \emph{Obtaining sharp peaks with large but finite mass:}
 The transition from steady state to runaway growth at a particular site is associated with divergence of the average mass
 at that site (see eq. \eqref{eqn:eq2b}, where  the mean
 mass $\langle m\rangle$ diverges as $\langle f[m]\rangle\rightarrow K_{sat}$).
 Thus, by tuning the system to be close to the phase boundary 
 between steady state and runaway growth (represented in fig. \ref{fig2} by the solid black line), the peak concentration of particles for any species can be made very high.
 (For instance, see fig. \ref{fig2}, where parameter combinations in the shaded region  
 result in a steady state in which the peak concentration of $B$ is greater than $2000$ particles at a site).
 Moreover, even though the $\langle f[m]\rangle$ profile is relatively smooth, the $\langle m\rangle$ 
 profile becomes very sharply peaked when the system is poised close to the phase boundary, leading to a sharply localized B-rich region. 

 \item  \emph{Tuning the locations of mass peaks:} The location of the peak of any mass profile (here 
 the $\langle m^{B}(x)\rangle$ profile) is determined by two length scales, $l_{c}$ and $l_{t}$, which govern the rise and fall of the  
 $\langle f[m^{B}(x)]\rangle$ and $\langle m^{B}(x)\rangle$ profiles near the cis and trans ends respectively. 
 For the two-species case, eq. \eqref{eqn:eq3} can be analyzed (see \cite{Himani2011}) to show that  
 (i) the length scale $l_{c}$ is small if the ratio $u/w_{A}$ is large: For high values of $u/w_{A}$,  most A particles undergo an $A\rightarrow B$ conversion before they can travel
into  the bulk, so that the concentration of $B$ particles rises steeply close to the source itself, leading to a small $l_{c}$
(ii) $l_{c}$ is large if the asymmetry factor $\gamma$ is high: As $\gamma$ increases, the proportion of B particles traveling in the anterograde direction rises, 
shifting the region of high B concentration in  this direction, thus leading to large $l_{c}$. (iii)
The length scale $l_{t}$, which governs the spatial variation of $\langle f[m^{B}(x)]\rangle$ near the trans end, only depends on the asymmetry factor $\gamma$, and is large when $\gamma$ is small:
The presence of the particle `sink' at the trans end lowers particle concentration in this region.
If there is extensive recycling or movement of particles in the retrograde direction (corresponding to $\gamma$ close to $1/2$), then the effect of the sink is `transmitted' in this direction, lowering the local mass  farther away    
 from the trans end, resulting in large $l_{t}$. The locations of the maxima of mass profiles in the 3-species model are also determined in a similar manner by the ratio of interconversion to fission rates and 
the degree of anterograde-retrograde asymmetry of particle movement. 
 \item \emph{Coarsegraining:}
 From eq. \eqref{eqn:eq3}, we can see that the mass profile remains unchanged under the scaling: $L\rightarrow \lambda L$,
  $u\rightarrow \lambda^{2}u$, $\left(\gamma-1/2\right)\rightarrow \lambda^{-1}\left(\gamma-1/2\right)$. This provides a 
  basis for coarsegraining the system, allowing us to reduce the model to a small number of `effective' lattice sites. 
However, for the coarsegraining procedure to be meaningful, this effective number should still be large enough that the
continuum approximation used in deriving \eqref{eqn:eq3} holds. 
 \end{enumerate}

 \subsection{Structure formation with different types of flux kernels $f[m]$}
\begin{wrapfigure}{r}{0.42\textwidth}
\vspace{-0.2cm}
\begin{center}
\includegraphics[width=0.38\textwidth]{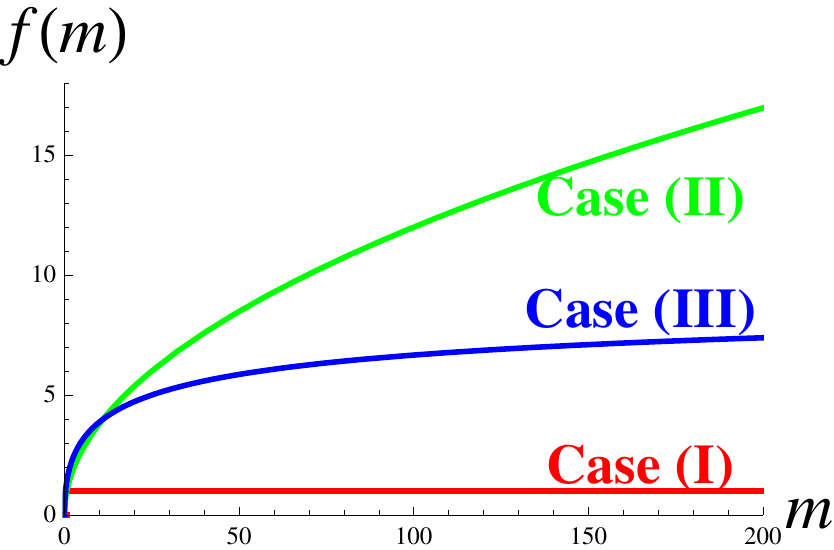}
\end{center}
\caption{A schematic of the three types of flux kernels $f[m]$ defined in eqs. \eqref{eqn:eq4a}-\eqref{eqn:eq4c}.}
\label{fig3}
\vspace{-1.2cm}
\end{wrapfigure}

 The above analysis suggests that the flux kernel $f[m^{Z}]$, which encapsulates how the vesicle fission and 
conversion rates depend on the mass of the parent aggregate, is a crucial determinant of structure formation. 
Below we present a detailed comparison of the structures obtained for three different kinds of $f[m^{z}]$
 (represented schematically in figure \ref{fig3}): \\ 

 \begin{enumerate}
\renewcommand{\theenumi}{\Roman{enumi}}
 \item Mass-independent rates: 
\begin{equation}
\label{eqn:eq4a}
f[m^{Z}]=\begin{cases}
0, & \text{if} \quad m^{Z}=0\\
1, & \text{if} \quad m^{Z}>0
\end{cases}
\tag{4a}
\end{equation}
\item Rate increasing with number $m_{i}^{Z}$ of $Z$ particles: 
\begin{equation}
\label{eqn:eq4b}
f[m^{Z}] \propto \sqrt{m^{Z}}
\tag{4b}
\end{equation}
\item Michelis-Menten type of rates:
 \begin{equation}
 \label{eqn:eq4c}
 f[m^{Z}]=\frac{K_{sat}\sqrt{m^{Z}}}{\sqrt{m^{Z}}+\sqrt{m_{sat}}} 
\tag{4c}
 \end{equation}
\end{enumerate}

\addtocounter{equation}{+1}

The above forms of $f[m^{Z}]$ emerge naturally for processes that are catalyzed by enzymes: If the number of enzyme 
molecules is much lower than the number of $A,B,C$ particles available for fission or conversion, then the reaction rate is 
enzyme-limited and roughly independent of $m^{A},m^{B}\ldots$ [case (I)]. If the enzyme concentration is in excess of the reactant particles, then the reaction is substrate-limited and the 
reaction rate is proportional to the number of $A,B,C$ particles available for reaction, leading to 
$f[m^{Z}]\propto [m^{Z}]^{\beta}$ where $\beta>0$ 
\footnote{$\sqrt{m^{Z}}$ rates emerge if we assume that the mass at any location in the 1D model is obtained by integrating 
over the 2D disc-like structure. A 2D structure with $m$ particles typically has $\sqrt{m}$ molecules at its perimeter. Assuming 
that only the perimeter molecules are available for reaction, we get reaction rates that are proportional to $\sqrt{m^{Z}}$}  
[case (II)]. For intermediate enzyme concentrations, we expect Michelis-Menten type of kinetics with the reaction rate increasing as 
$\sqrt{m^{Z}}$  for small $m^{Z}$ and saturating to a constant value at large $m^{Z}$ [case (III)].

\begin{figure}[h!]
\centering
   \renewcommand{\thesubfigure}{a}
\subfloat[]{\label{fig4a1}
  \includegraphics[width=4.9cm,height=3.528cm]{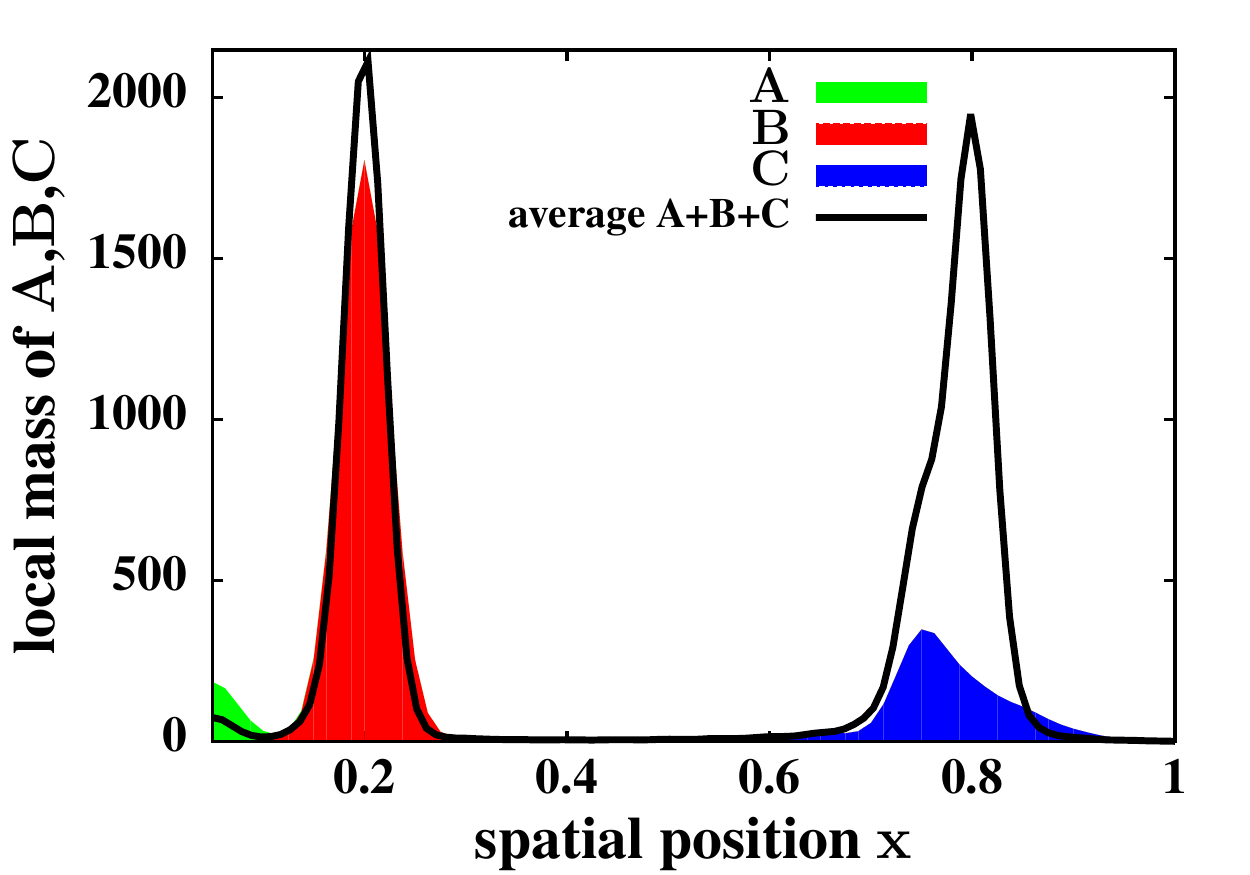}
  }
   \renewcommand{\thesubfigure}{c}
  \subfloat[]{\label{fig4b1}
  \includegraphics[width=4.9cm,height=3.528cm]{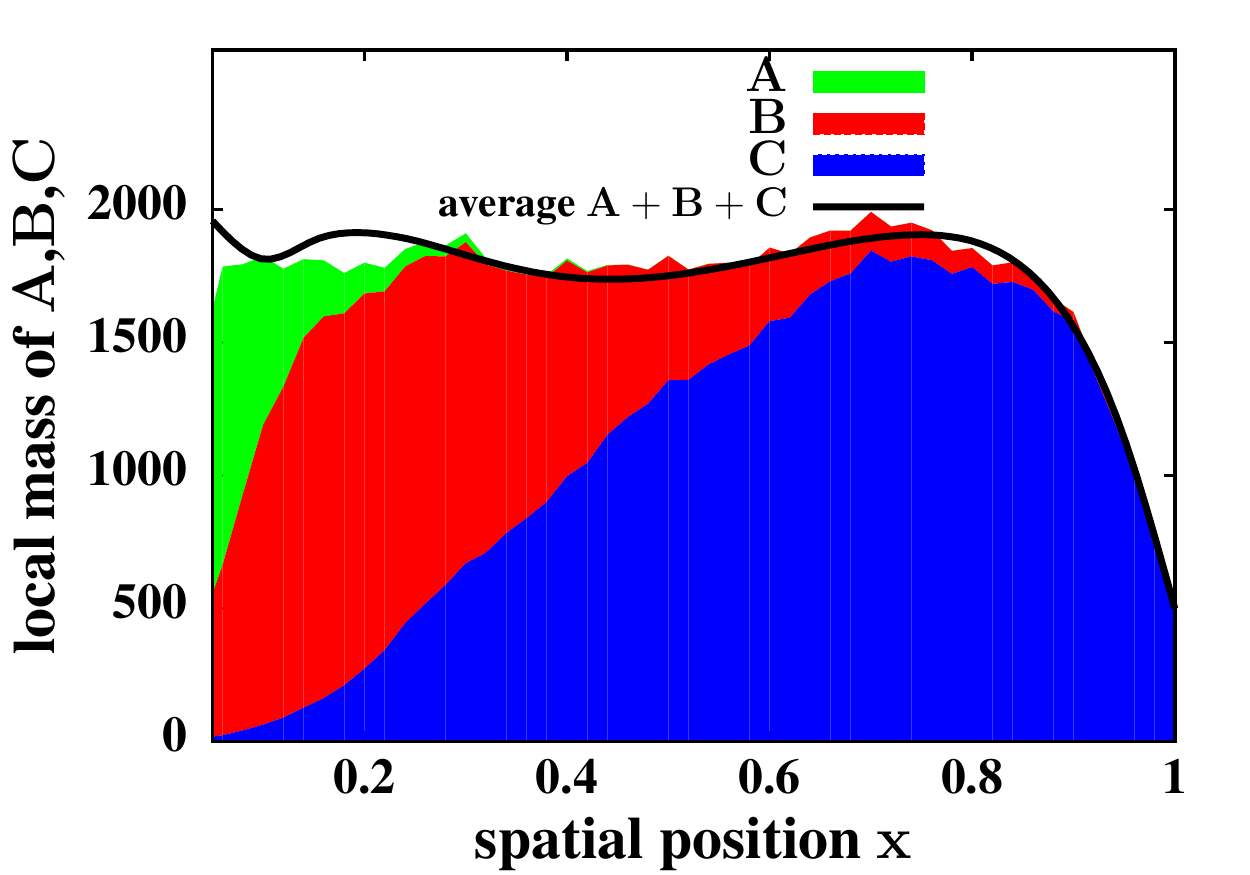}
  }
     \renewcommand{\thesubfigure}{e}
\subfloat[]{\label{fig4c1}
  \includegraphics[width=4.9cm,height=3.528cm]{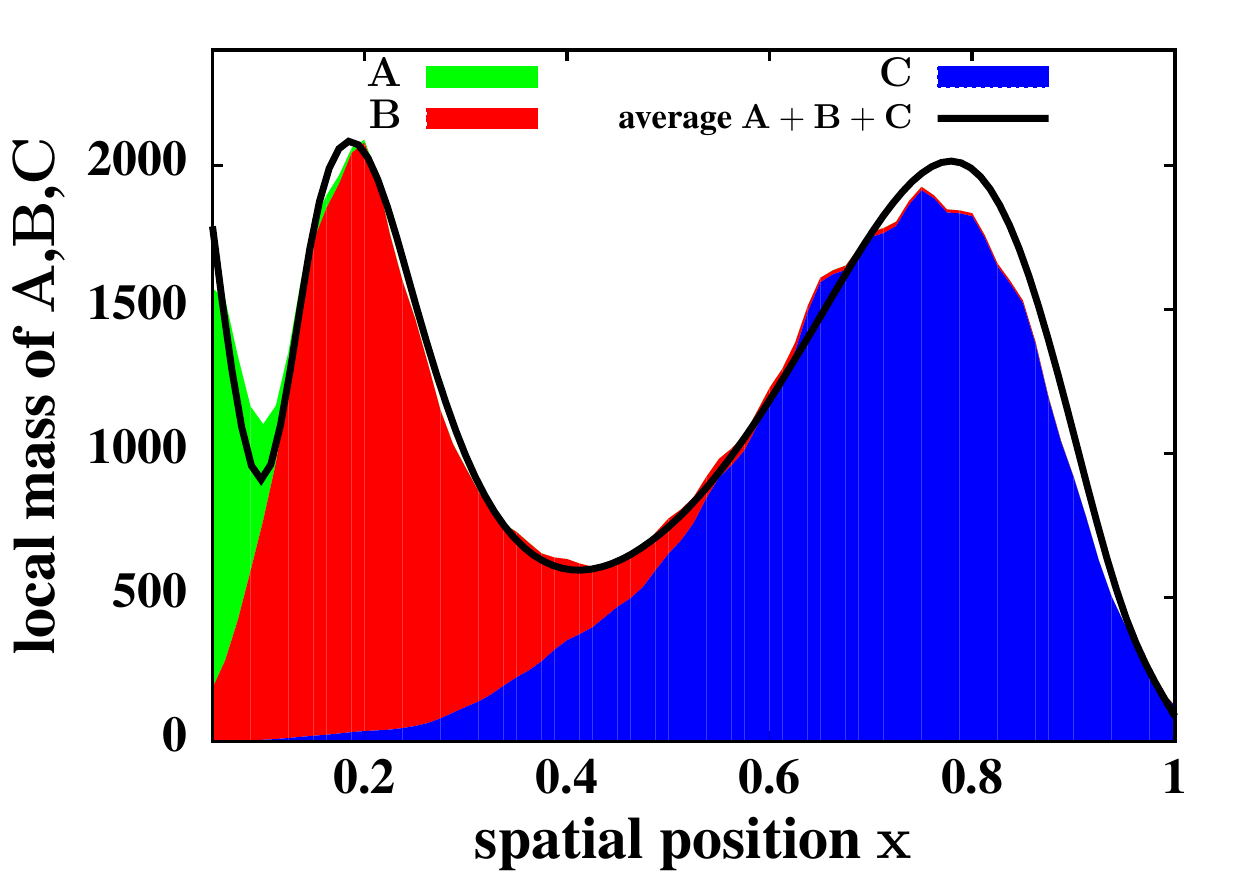}
  }\\
     \renewcommand{\thesubfigure}{b}
  \subfloat[]{\label{fig4a2}
  \includegraphics[width=4.9cm,height=3.528cm]{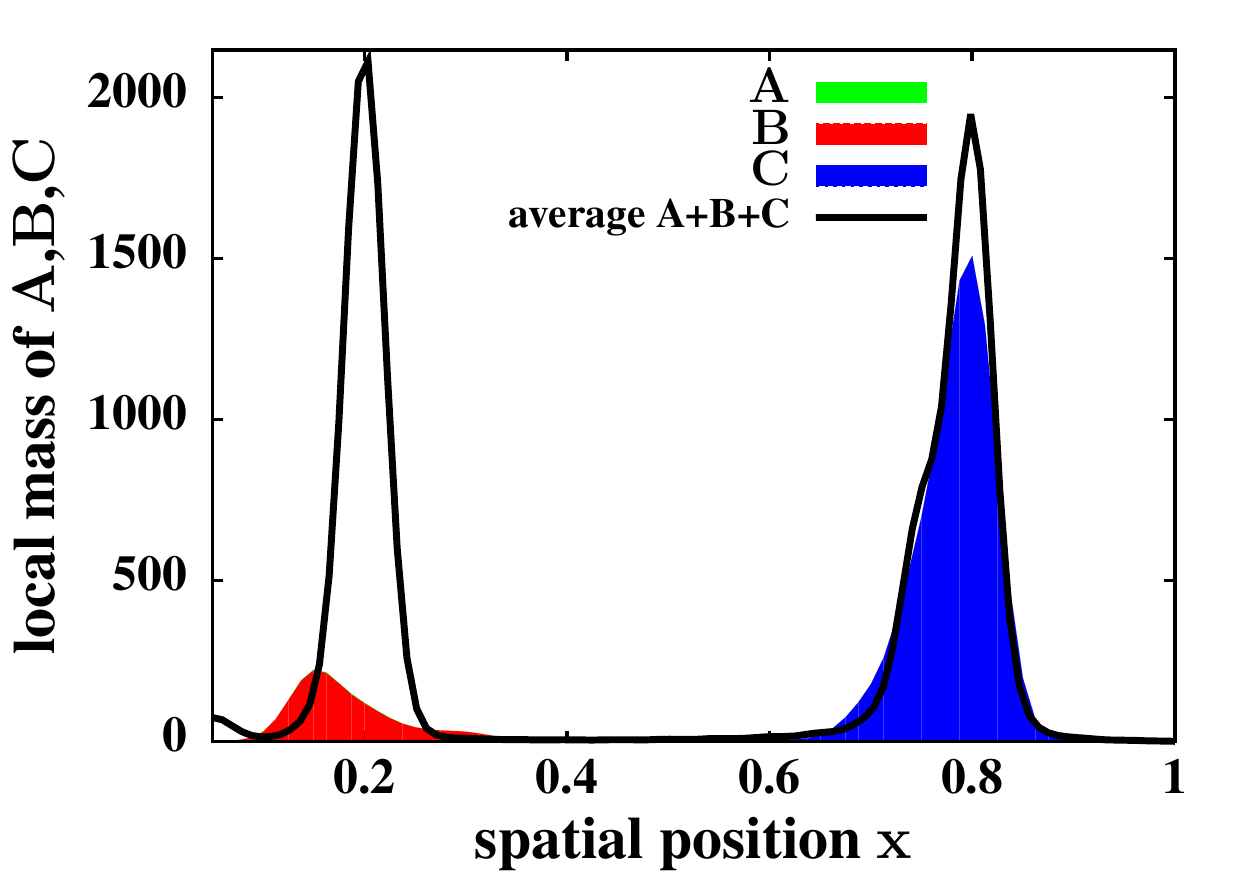}
  }
     \renewcommand{\thesubfigure}{d}
\subfloat[]{\label{fig4b2}
  \includegraphics[width=4.9cm,height=3.528cm]{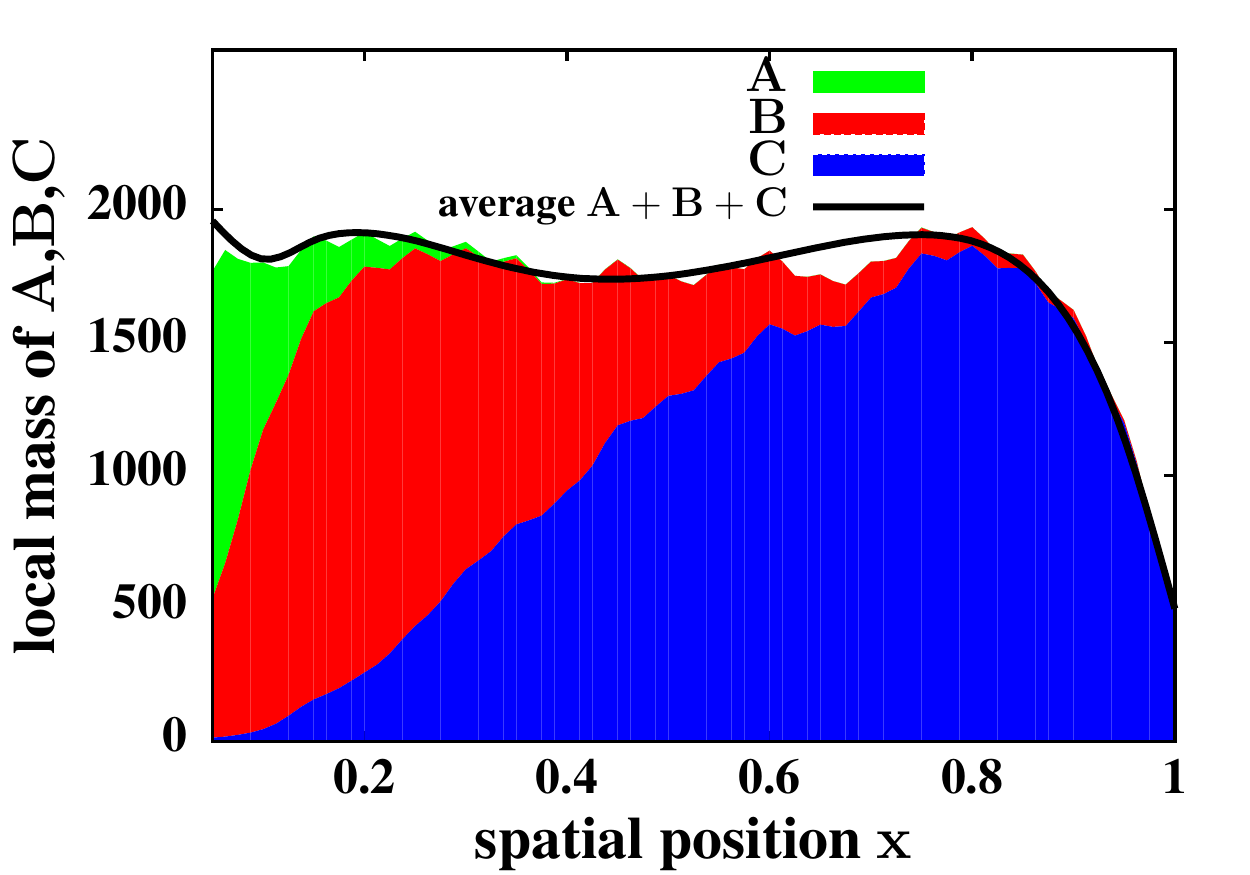}
  }
     \renewcommand{\thesubfigure}{f}
\subfloat[]{\label{fig4c2}
  \includegraphics[width=4.9cm,height=3.528cm]{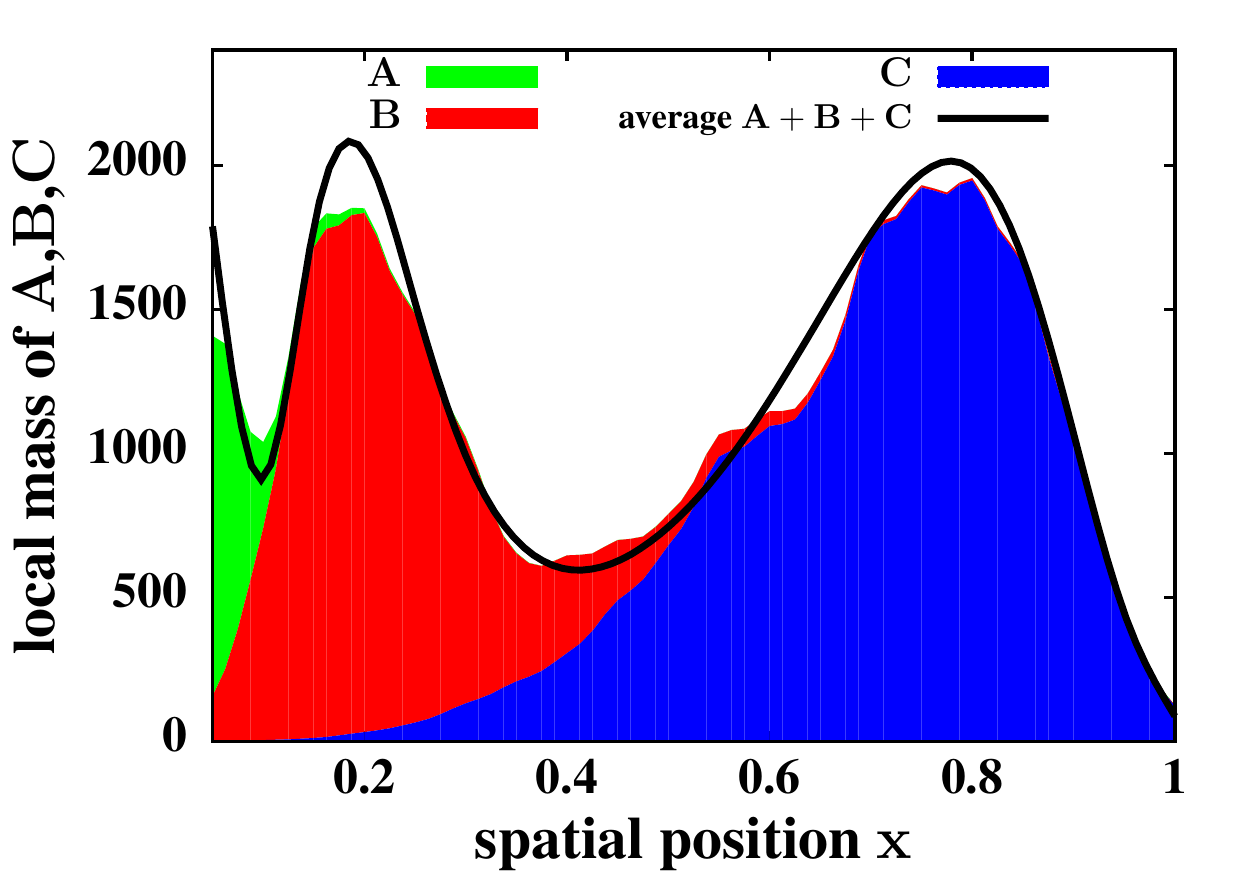}
  }
\caption{Typical snapshots of the system with pure VT for the three types of $f[m]$ (at two different time instants). (a)-(b): For type I $f[m]$, B, C compartments are 
well-separated but fluctuate strongly about the average mass profiles (solid lines).  (c)-(d): For type II $f[m]$, compartments cannot be resolved into morphologically distinct 
structures but are very stable in time.  (e)-(f): For type III $f[m]$, compartments are both morphologically distinct and reasonably stable in time.
Parameters for (a)-(b): $a=1$, $D=0$, $w_{A}=0.09368$, $w_{B}=0.03346$, $w_{C}=0.04175$, $\gamma_{A}=0.5$, $\gamma_{B}=0.6$, 
 $\gamma_{C}=0.66$, $u=0.0086$, $v=0.00149$, $m_{sat}=200$, $K_{sat}=14.14$; (c)-(d): $a=1$, $D=0$, $w_{A}=0.03448$, $w_{B}=0.01207$, $w_{C}=0.01379$, $\gamma_{A}=0.5$, $\gamma_{B}=0.6$, 
 $\gamma_{C}=0.66$, $u=0.0031$, $v=0.00053$, $m_{sat}=200$, $K_{sat}=14.14$; (e)-(f): $a=1$, $D=0$, $w_{A}=0.125$, $w_{B}=0.04375$, $w_{C}=0.05462$, $\gamma_{A}=0.5$, $\gamma_{B}=0.6$, 
 $\gamma_{C}=0.66$, $u=0.01125$, $v=0.00194$, $m_{sat}=200$, $K_{sat}=14.14$.}
\label{fig4}
\end{figure}

\paragraph*{}
Note that cases (I) and (III) admit runaway growth, as the function $f[m^{Z}]$ is bounded for large $m^{Z}$. Thus, in order to 
generate finite aggregates with constant mass, we must choose parameters that result in steady state [region in red in fig. \ref{fig2}]. 
As discussed in sec. \ref{sec:sec2.1}, we can solve for $\langle f[m^{Z}(x)]\rangle$ in steady state, and approximately derive the mass profiles 
$\langle m^{Z}(x)\rangle$ from these solutions, using the following approximations:
\begin{subequations}
 \begin{equation}
  \langle m^{Z}\rangle \sim \frac{\langle f[m^{Z}]\rangle}{1-\langle f[m^{Z}]\rangle} \qquad  \qquad \text{type I }f[m]
 \label{eqn:eq7a}
  \end{equation}
\begin{equation}
 \langle m^{Z}\rangle \sim \langle f[m^{Z}]\rangle^{2}  \quad \qquad \qquad  \qquad \text{type II }f[m]
 \label{eqn:eq7b}
 \end{equation}
 \begin{equation}
 \langle m^{Z}\rangle\sim \left( \frac{\sqrt{m_{sat}}\langle f[m^{Z}]\rangle}{K_{sat}-\langle f[m^{Z}]\rangle}\right)^{2}  \qquad \text{type III }f[m]
 \label{eqn:eq7c}
 \end{equation}
  \label{eqn:eq7}
\end{subequations}
Equation \eqref{eqn:eq7a} has been derived explicitly for the single-species model without interconversion 
in \cite{Levine2005} and appears to hold in the presence of interconversion as well in the limit $f[m]\rightarrow 1$.
Equations \eqref{eqn:eq7b}-\eqref{eqn:eq7c} can be derived using the approximation $\langle f[m]\rangle \sim f[\langle m\rangle]$, 
which is reasonable for $\langle m\rangle \gg 1$.
Based on the approximations \eqref{eqn:eq7a}-\eqref{eqn:eq7c}
and numerical simulations, we now compare in detail various properties of the structures generated for the 
three types of $f[m]$.
In particular, we consider how the function $f[m]$ governs (i) where compartments form,  i.e., the spatial location of the maxima of mass profiles 
(ii) how localized compartments are, i.e., the width of the maxima (iii) whether compartments are temporally stable,  i.e., the fluctuations of mass profiles about the average, and  
(iv) the robustness of structures to variations in rates, i.e., the relative change in mass profiles due to changes in parameters.
Some of these features can also be observed in typical snapshots of structures (fig. \ref{fig4}), where the abundances of $A$,$B$,$C$ particles at each location
are represented by the heights of the green, red, blue columns at that location.

\paragraph*{Location of maxima:}
The spatial locations of the maxima of the $\langle m^{A}\rangle$, $\langle m^{B}\rangle$, $\langle m^{C}\rangle, \ldots$  profiles are roughly independent of the form 
of $f[m]$ and are determined primarily by the rates of the elementary processes in the model. To see this, note that for each of the 
three forms (I)-(III), the derivative $d\langle m^{Z}(x)\rangle/dx$ can be zero if and only if the derivative 
$d\langle f[ m^{Z}(x)]\rangle/dx$ is zero (from eq. \eqref{eqn:eq7}),
implying that the maximum of the $\langle m^{Z}(x)\rangle$ and $\langle f[m^{Z}(x)]\rangle$  profiles 
must coincide. Since the spatial profiles $\langle f[m^{Z}(x)]\rangle$ depend only on the rates of various processes (see eq. \eqref{eqn:eq3}), and not on the form of $f[m]$, it follows that the  
corresponding $\langle m^{Z}(x)\rangle$ profiles and the positions of their maxima must also be independent of the form of $f[m]$.

\paragraph*{Width of maxima:}
In order to obtain non-overlapping compartments, i.e., well-separated peaks in the total mass profile, the width of the maxima 
must be much smaller than the distance between adjacent maxima.  
Sharp peaks (with small width) can be obtained only for flux kernels $f[m]$ of types (I) and (III), that 
saturate at large $m$.
As evident from eqs. \eqref{eqn:eq7a} and \eqref{eqn:eq7c}, if the average vesicular fluxes at the peaks locations are close to their saturation value, then the average mass at the 
peaks is very high. Moreover, in this saturation regime, gentle gradients in the vesicular fluxes, i.e., in $\langle f[m^{Z}(x)]\rangle$ profiles, can result in very sharp peaks in the $\langle m^{Z}(x)\rangle$ profiles, 
thus, leading to sharply localized $A$-rich, $B$-rich and $C$-rich regions  
(see also sec. \ref{sec:sec2.1}). 
For $f[m]$ of type (II), there is no such saturation regime; the average mass profiles $\langle m^{Z}(x)\rangle$ essentially mirror the $\langle f[m^{Z}(x)]\rangle$ profiles (eq. \eqref{eqn:eq7b}), and have 
maxima that are wide and difficult to resolve.
Average mass profiles for the three types of $f[m]$ are depicted in  figs. \ref{fig4}\subref{fig4a1}-\ref{fig4}\subref{fig4c1} as solid lines,
 for parameters chosen so as to obtain spatially resolvable peaks (which was not possible in case II).
 Note that peaks are much sharper in case I [figs. \ref{fig4}\subref{fig4a1},\ref{fig4}\subref{fig4a2}] than in case III
[figs. \ref{fig4}\subref{fig4c1},\ref{fig4}\subref{fig4c2}].

 

\begin{figure}[h!]
\begin{center}
\subfloat[]{\label{fig5a}
  \includegraphics[width=8.5cm,height=6.12cm]{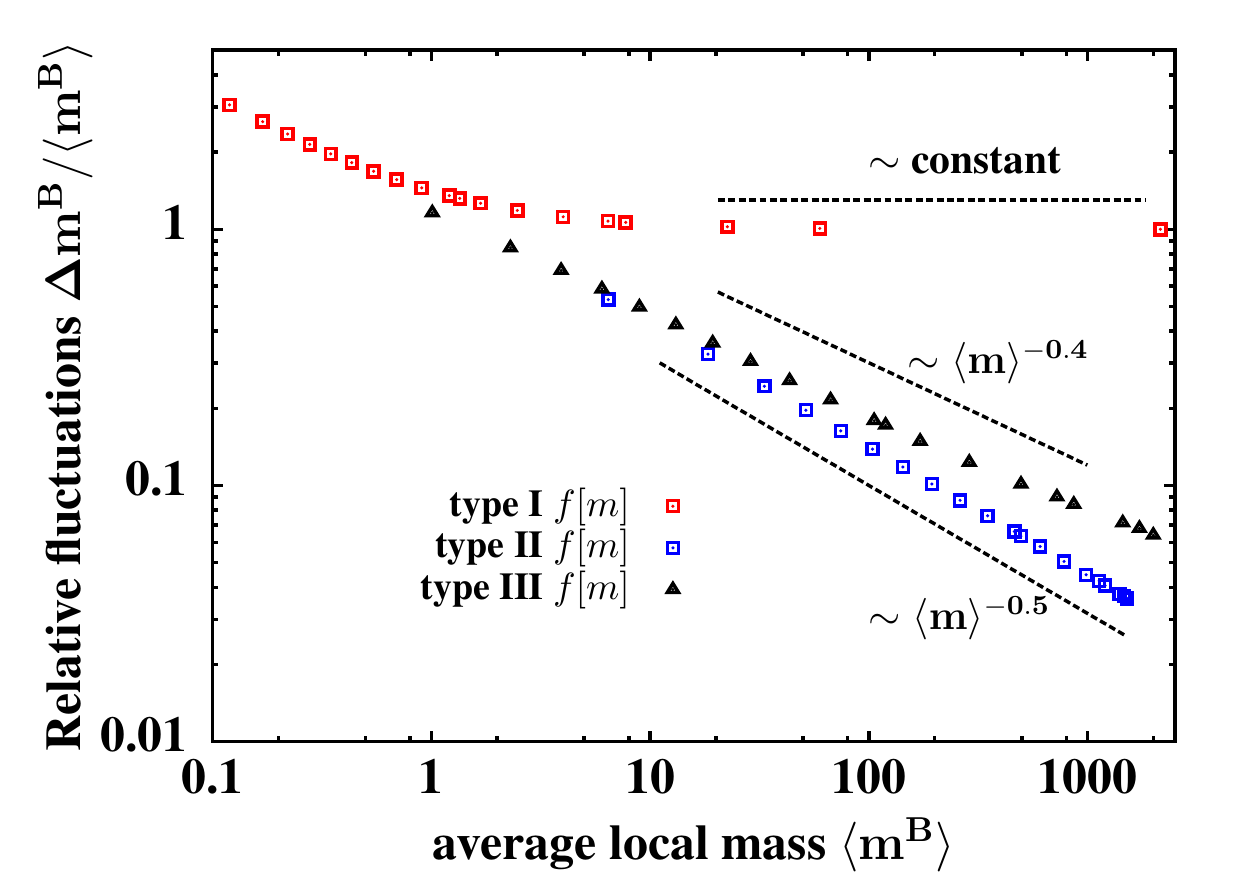}
  } 
\subfloat[]{\label{fig5b}
  \includegraphics[width=8.5cm,height=6.12cm]{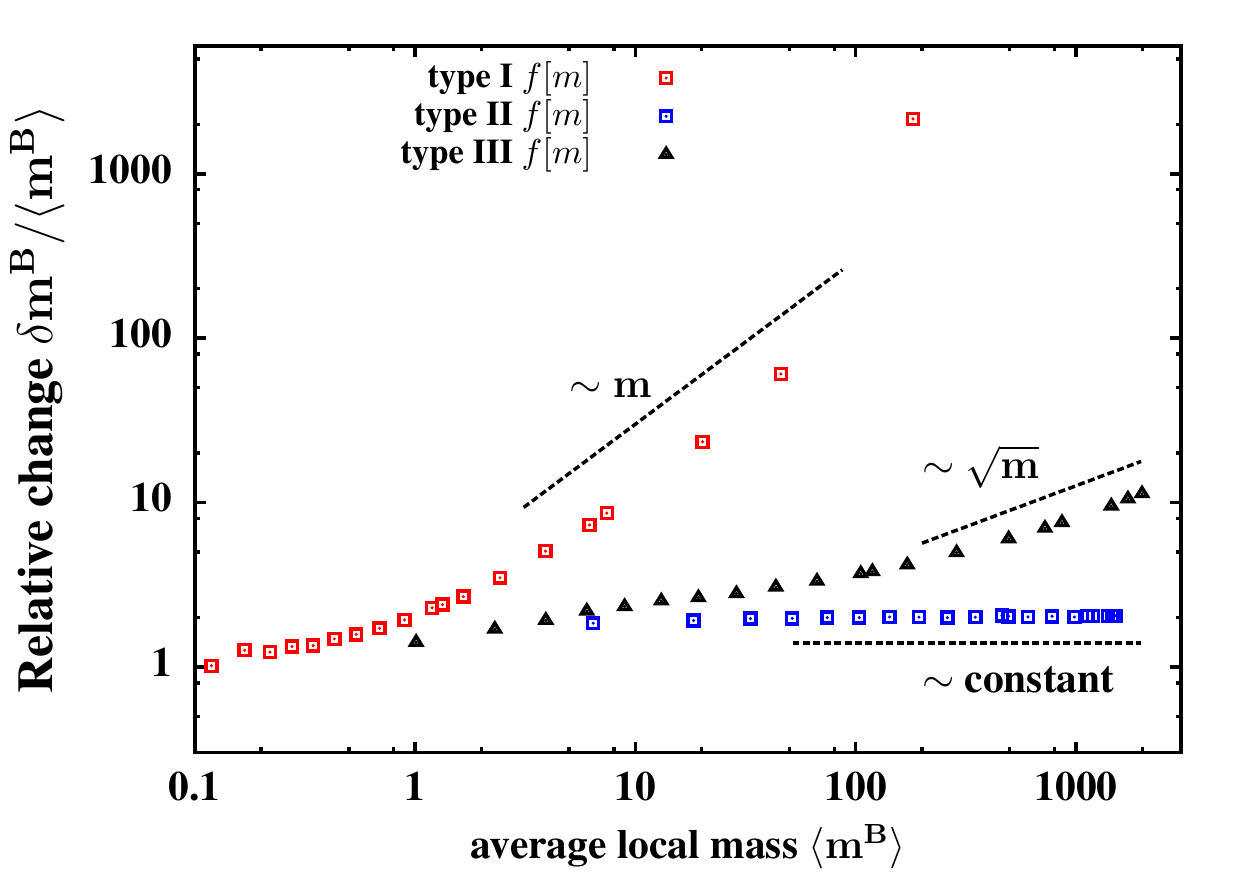}
  }
\end{center}
\vspace{-0.5cm}
\caption{(a) Log-log plot of relative fluctuations $\Delta m^{B}/\langle m^{B}\rangle$ vs. the \emph{local}
mass $\langle m^{B}\rangle$  of $B$ particles. For type I $f[m]$, 
$\Delta m^{B}/\langle m^{B}\rangle\sim 1$, indicating giant fluctuations about the average mass profile. 
For $f[m]$ of type II and III, $\Delta m^{B}/\langle m^{B}\rangle$ decreases 
with $\langle m^{B}\rangle$, implying that large aggregates are relatively stable.
(b) Log-log plot of the relative change $\delta  m^{B}/\langle m^{B}\rangle$ at a location  with average local mass $\langle m^{B}\rangle$ of B, due to change in influx. Y-axis is scaled by    
$\delta a/a$, the increase in influx, where $\delta a/a=0.005$ for $f[m]$ of type I and $\delta a/a=0.05$ for type
II and III $f[m]$. The relative change is independent of $\langle m^{B}\rangle$ for case II, but increases with $\langle m^{B}\rangle$
for cases I and III. Parameters same as for fig. \ref{fig4}.
}
\label{fig5}
\end{figure}

\paragraph*{Stability of mass profiles:} 
To estimate the extent to which mass profiles vary over time, we monitor the rms fluctuations 
$\Delta m^{Z}$ about the average local mass $\langle m^{Z} \rangle$ of each species at each site.
Figure \ref{fig5}\subref{fig5a} shows the relative fluctuations $\Delta m^{B} /\langle m^{B}\rangle$ vs. $\langle m^{B}\rangle$ for the three cases. For mass-independent $f[m]$ (type I), $\Delta m^{B} /\langle m^{B}\rangle \sim 1$
for large $\langle m^{B}\rangle$, so that  mass profiles exhibit giant fluctuations about the average (figs. \ref{fig4}\subref{fig4a1}-\ref{fig4}\subref{fig4a2}). For $f[m]$ of type II and III, where the flux increases with $m$,
the relative fluctuations $\Delta m^{B} /\langle m^{B}\rangle$ become smaller as  $\langle m^{B}\rangle$ increases, so that large aggregates are fairly stable in time (see figs. 
 \ref{fig4}\subref{fig4b1}-\ref{fig4}\subref{fig4c2}).
 Thus, vesicular fluxes must be sensitive to the mass of the 
parent aggregate to ensure a a stabilizing negative feedback-- when
 the aggregate becomes too large, the number of particles breaking off from it increases and vice versa, thus tending to
 restore the aggregate to its average size. 

 \paragraph*{Sensitivity of mass profiles to small variations in rates:}
Changes in model parameters can alter the 
  peak locations of $\langle m^{Z}(x)\rangle$ profiles as well as peak concentrations. However, in the regime of small 
 interconversion rates ($u,v\ll w_{A},w_{B}$) that we consider here, the locations are relatively insensitive to changes in 
 parameters, and 
 perturbations primarily alter the peak concentrations of some or all $\langle m^{Z}\rangle$ profiles.
\paragraph*{}
Below we consider how mass profiles respond to a change in influx,  
by measuring $[\delta m^{Z}/\langle m^{Z}\rangle]/[\delta a/a]$ which is the ratio of the 
relative change $\delta m^{Z}/\langle m^{Z}\rangle$ that occurs at a location with average mass $\langle m^{Z}\rangle$, to the  
corresponding change $\delta a/a$ in influx. From eqs. \eqref{eqn:eq3} and \eqref{eqn:eq7}, it follows that for $\langle m^{Z}\rangle\gg 1$ and 
 $\delta a/a\rightarrow 0$, we expect:
  $\delta m^{Z}/\langle m^{Z}\rangle \propto \langle m^{Z}\rangle(\delta a/a)$ for $f[m]$ of type I;
$\delta m^{Z}/\langle m^{Z}\rangle \propto 2(\delta a/a)$ for $f[m]$ of type II;
  $\delta m^{Z}/\langle m^{Z}\rangle \propto 2\left[1+\sqrt{\langle m^{Z}\rangle/m_{sat}}\right](\delta a/a)$ for $f[m]$ of type III. 
These expectations are confirmed qualitatively by numerics (fig. \ref{fig5}\subref{fig5b}), with  
the strongest response to change in influx observed for type (I) $f[m]$, followed by type (III), and the 
weakest for $f[m]$ of type (II). Furthermore, with type (II) $f[m]$, the relative change in mass is the same in all regions, 
whereas with $f[m]$ of type (I) and (III), the relative change is more in regions where the local mass is already high.
Thus, with type (I) and (III) $f[m]$, the mass profile becomes more/less sharply peaked as influx 
increases/decreases (see also inset of fig. 5 of main paper). 
Note that for these types of $f[m]$, a large increase  in influx  can drive  
 an instability leading to runaway growth-- an undesirable feature which is eliminated if there is sub-cisternal movement  
at a small rate (sec. \ref{sec:sec3}).
\paragraph*{}
This analysis also sheds light on the degree of fine tuning required to generate compartments. Since the mass profiles are most (least) sensitive to small changes in parameters when $f[m]$ is of type I (type II), it follows that the 
maximum (minimum) degree of fine tuning is required in this case. By the same token, an intermediate degree of fine tuning is required for structure formation with type III $f[m]$. 
 
 \subsection{Pure VT model with higher number of species}
 \begin{figure}[h]
\subfloat[] {
\centering
\includegraphics[width=7cm,height=4.8cm]{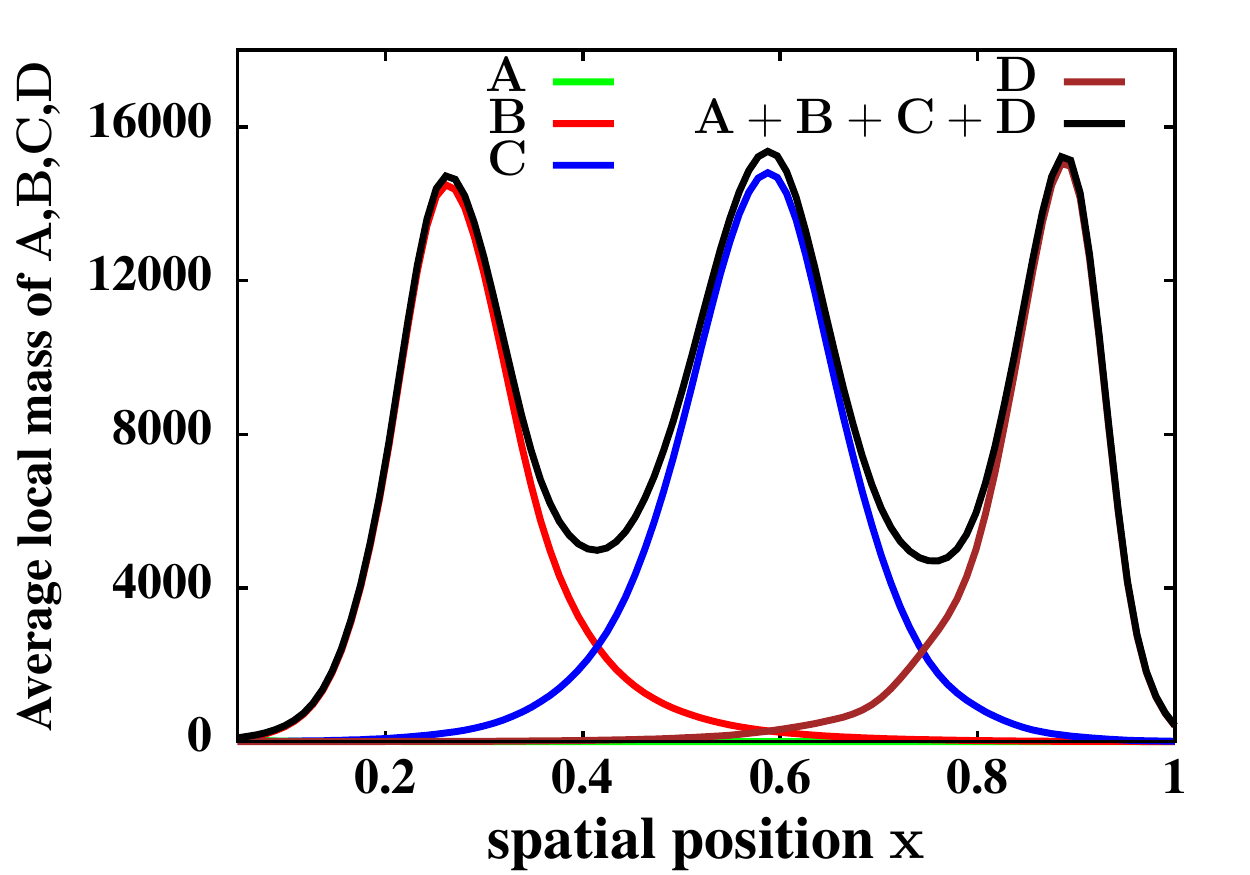}
\label{fig6a}}
\subfloat[]{
\includegraphics[width=7cm,height=4.8cm]{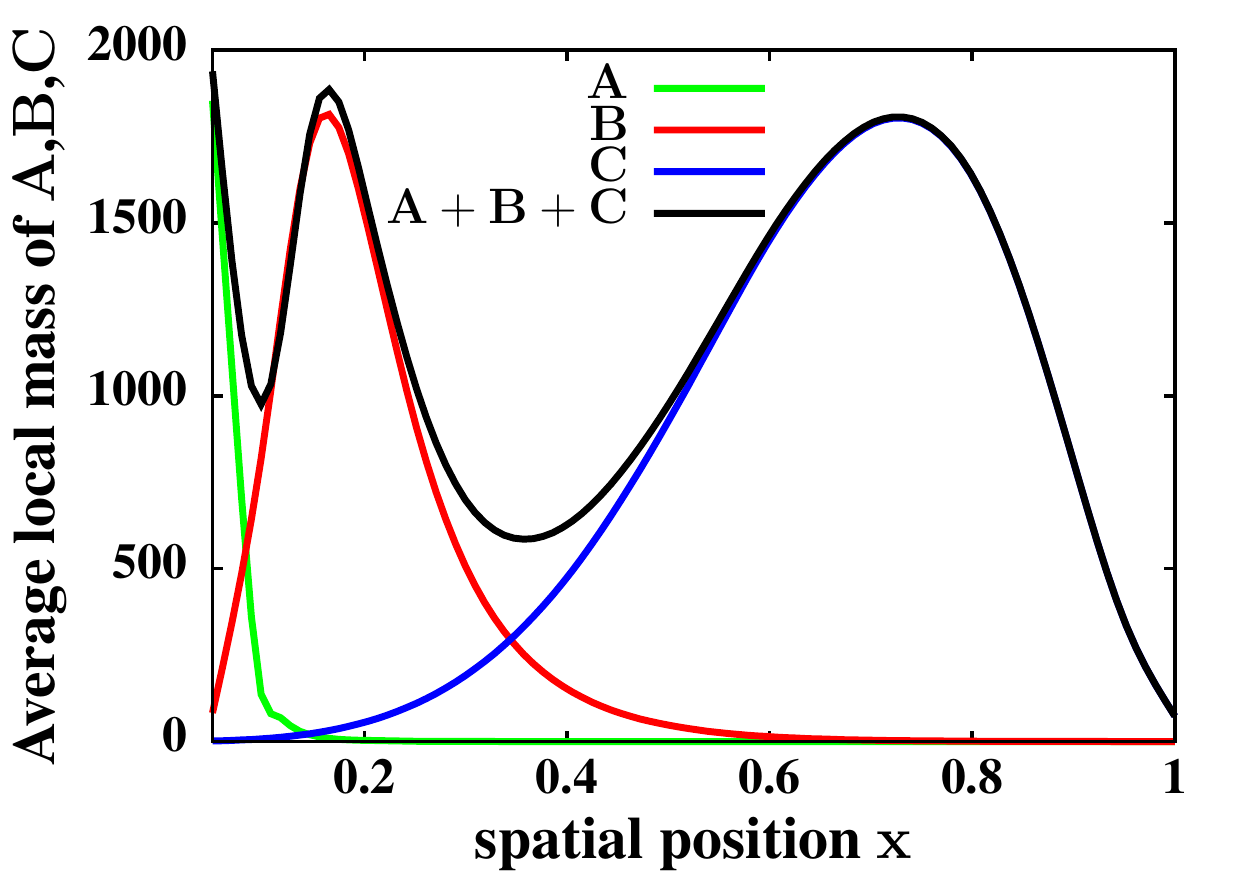}
\label{fig6b}}
\caption{Two variants of the basic model.
(a) Four-species model with A,B,C,D particles, sequential $A\rightarrow B\rightarrow C\rightarrow D$ conversion
and MM flux kernel. System self-organizes into distinct $B$, $C$, $D$ compartments. Parameters: $a=1$, $D=0$, $w_{A}=0.3419$, $w_{B}=0.12863$, $w_{C}=0.00504$, $w_{D}=0.00018$ $\gamma_{A}=0.5$, $\gamma_{B}=0.5$, 
 $\gamma_{C}=0.54$, $\gamma_{D}=0.75$, $u=0.02564$, $v=0.000079$, $r=0.0000071$\protect\footnotemark, $m_{sat}=100$, $K_{sat}=10$.
(b) Average mass profiles in the pure VT model with homotypic fusion of A and B particles. Note the similarity 
with figs. \ref{fig4}\protect\subref{fig4c1}, \ref{fig4}\protect\subref{fig4c2} where there is no homotypic fusion.
Parameters: $a=1$, $D=0$, $w_{A}=0.15094$, $w_{B}=0.03094$, $w_{C}=0.03472$, $\gamma_{A}=0.5$, $\gamma_{B}=0.63$, 
 $\gamma_{C}=0.66$, $u=0.00875$, $v=0.00146$, $m_{sat}=200$, $K_{sat}=14.14$, $m^{\prime}_{sat}=100$, $K^{\prime}_{sat}=10$.}
\label{fig6}
\end{figure}

The pure VT model can be generalized to include more species of particles; here we consider a 4-species version
with $A\rightarrow B\rightarrow C\rightarrow D$ sequential interconversion. 
 Figure \ref{fig6}\subref{fig6a} shows the average mass profiles for a set of parameters 
 chosen such that the maxima of the $B$ , $C$ and $D$ profiles are sufficiently well separated. The maximum of each mass profile can be made sharp
by tuning the maximum of the corresponding $\langle f[m^{Z}]\rangle$ profile to be close to the saturation
value $K_{sat}$, so that the total mass profile has three distinct 
peaks corresponding to morphologically separate $B$, $C$ and $D$ compartments.
The model can be extended in a similar fashion to include more number of species, and thus, obtain a higher number 
of  compartments. 

\footnotetext{$r$ is the $C\rightarrow D$ conversion rate.}

 \subsection{Effect of homotypic fusion}
 We use numerical simulations to study the effect of introducing homotypic fusion in the pure VT model. Homotypic fusion 
 refers to the tendency of an A (or B) vesicle to fuse with a neighboring aggregate with a rate that increases with the 
 number of A (or B) vesicles in the target aggregate. To implement homotypic fusion in our model, we 
assume that an A particle at site $i$ to hops to its (forward) neighbor $j$ with a rate given by:
 \begin{equation}
  w_{A}\gamma_{A} \left(\frac{K_{sat}\sqrt{m^{A}_{i}}}{\sqrt{m^{A}_{i}}+\sqrt{m_{sat}}}\right)\left(0.2+0.8
 \frac{K'_{sat}\sqrt{m^{A}_{j}}}{\sqrt{m^{A}_{j}}+\sqrt{m^{\prime}_{sat}}} \right)
 \end{equation}
 The first bracketed term represents enzyme-mediated fission from site $i$ with MM kinetics. The second term represents homotypic fusion 
of the fissioned A particle with the mass at site $j$ at a rate that increases with the mass of A particles  
present at the site, but saturates to a constant value beyond a certain mass $m^{\prime}_{sat}$. The hopping rates 
for B particles can be written in a similar fashion. 
\paragraph*{}
Figure \ref{fig6}\subref{fig6b} depicts the mass profiles for parameters chosen such that both the fission and fusion terms  are operating
within the saturation regime. Under this condition, structures generated in the pure VT model with homotypic fusion of $A$, $B$ 
 (and all others excluding the final product) species,  are qualitatively similar 
to those generated without homotypic fusion [see figs. \ref{fig4}\subref{fig4c1}, \ref{fig4}\subref{fig4c2}]. Thus, in our model, homotypic fusion is not crucial for maintaining compartment identity.
 
 \section{VT-dominated case: Effect of (sub-)cisternal movement}
\label{sec:sec3}
\begin{wrapfigure}{r}{0.48\textwidth}
\vspace{-0.8cm}
\begin{center}
\includegraphics[width=0.47\textwidth]{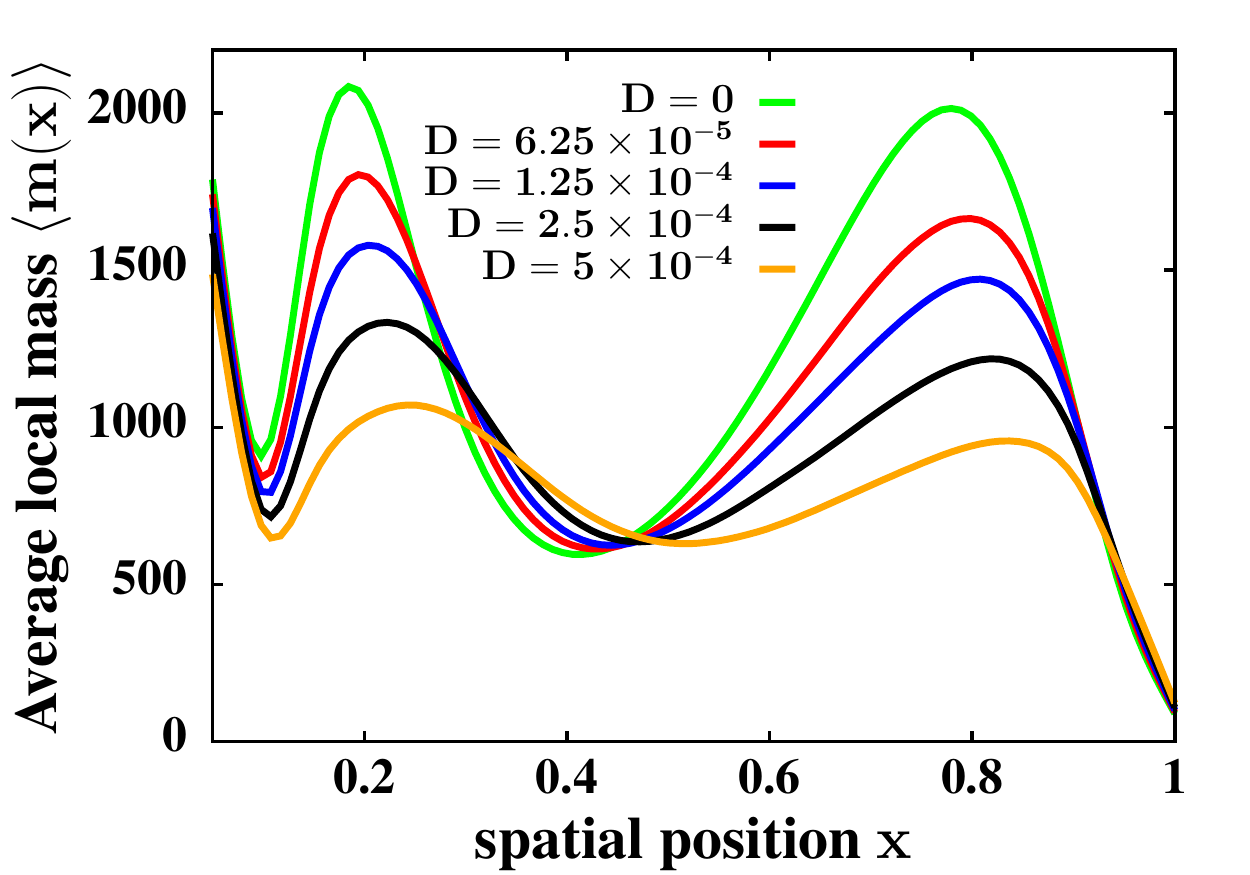}
\end{center}
\caption{Average mass profiles $\langle m(x)\rangle$ for different values of sub-cisternal movement rate $D$ (cisternal fraction $\alpha=0.3$ and other parameters same as in figs.
\ref{fig4}\protect\subref{fig4c1}, \ref{fig4}\protect\subref{fig4c2}). 
The  mass profiles become less peaked with increasing $D$, making it difficult to resolve adjacent maxima.}
\label{fig8}
\vspace{-0.9cm}
\end{wrapfigure}
\paragraph*{}
We consider a scenario where in addition to vesicular transport, a finite fraction $\alpha$ of an aggregate is allowed to break off and move forward with rate $D$. This is akin to the Cisternal Progenitor model of \cite{Pfeffer1}, and is referred to the 
 VT-dominated limit in the main paper. Breakage and movement of cisternal fragments at a non-zero rate $D$ 
 is sufficient to ensure that there is no runaway growth, even for very large influx rates $a$. To 
illustrate this point we analyze a very simple case where the system consists of a single site ($L=1$), and where there is 
no interconversion. The average mass on the site evolves as:
\begin{equation}
\frac{\partial \langle m^{A}(t)\rangle}{\partial t} = a-w_{A}\langle f[m^{A}] \rangle - \alpha D\langle m^{A} \rangle
\end{equation}
where $f[m^{A}]$ is of type III [eq. \eqref{eqn:eq4c}].
\paragraph*{}
For $a/p>K_{sat}$ and $D=0$, the incoming and outgoing flux at the site necessarily fail to balance, as $\langle f[m^{A}] \rangle$ cannot exceed $K_{sat}$, 
leading to runaway growth of mass.
Now consider the case with non-zero $D$.  As $\langle m^{A}\rangle$ increases, the average 
outgoing  flux (which includes the term $\alpha D\langle m^{A}\rangle$) also increases, until it balances the incoming flux $a$.
At this point, $d\langle m^{A}\rangle /dt$  must become 
zero, ensuring that  $\langle m^{A}\rangle$ is constant at long times. Thus, as long as the outgoing flux has a component that 
\emph{keeps increasing with the mass}  $m$ at the site, there can be no runaway growth.
\paragraph*{}
In order to preserve the well-separated compartments that emerge in the pure VT model, the rate $D$  must be smaller than the single particle fission rates. As $D$ increases, the peaks in the mass profile become shallower, until adjacent 
peaks can no longer be resolved (fig. \ref{fig8}). Thus, we consider VT-dominated scenarios where cisternal movement acts as a secondary track to vesicular transport.
 This eliminates runaway
growth while maintaining sharp peaks in the mass profile (see also figs. 2(c) and 2($c^{\prime}$) of the main paper).

  \begin{figure*}[t]
\centering
   \renewcommand{\thesubfigure}{a}
\subfloat[]{\label{fig_agg1}
  \includegraphics[width=7.3cm,height=5.0cm]{snap_MM_1}
  }
  \renewcommand{\thesubfigure}{b}
\subfloat[]{\label{fig_agg2}
  \includegraphics[width=7.3cm,height=5.0cm]{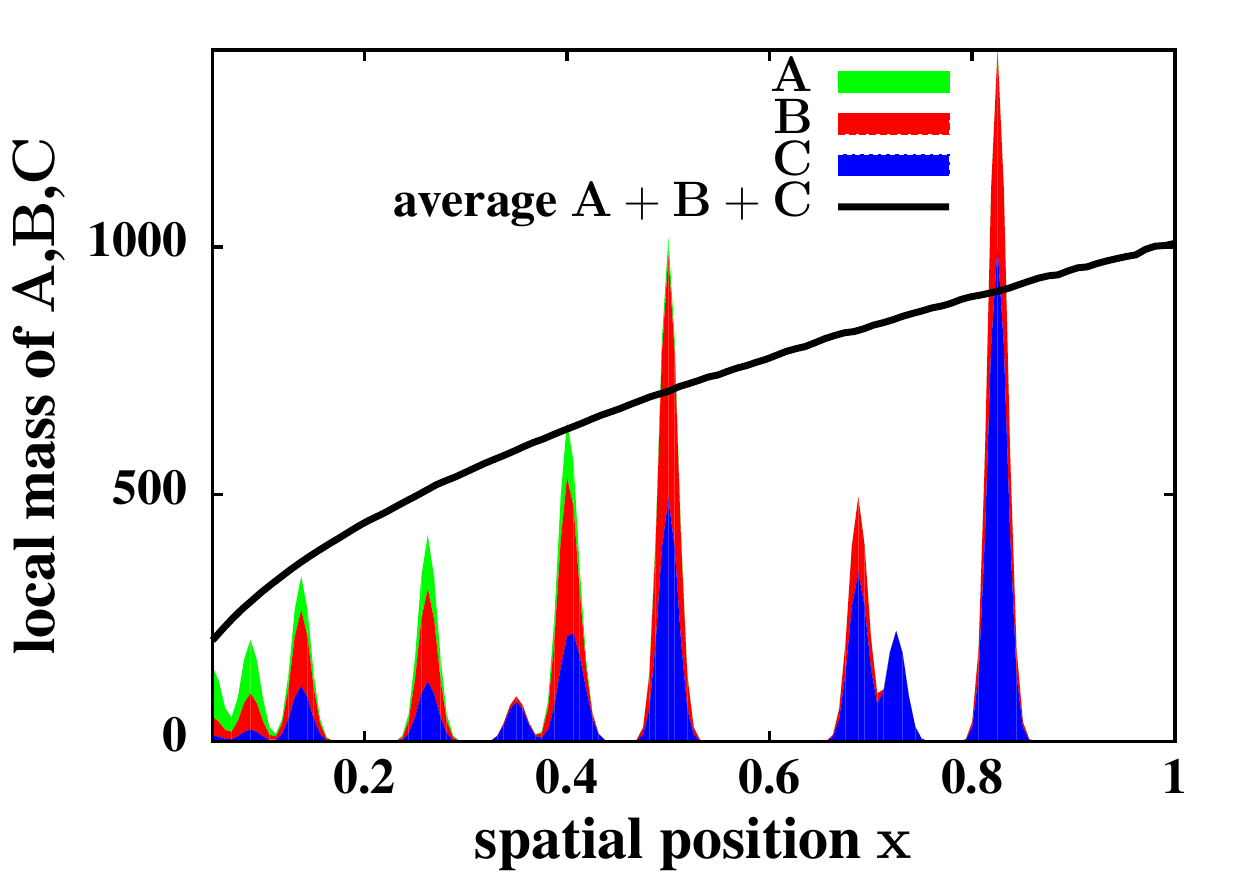}      
  }  
  \\
 \renewcommand{\thesubfigure}{c}
  \subfloat[]{\label{fig_agg3}
  \includegraphics[width=7.3cm,height=5.0cm]{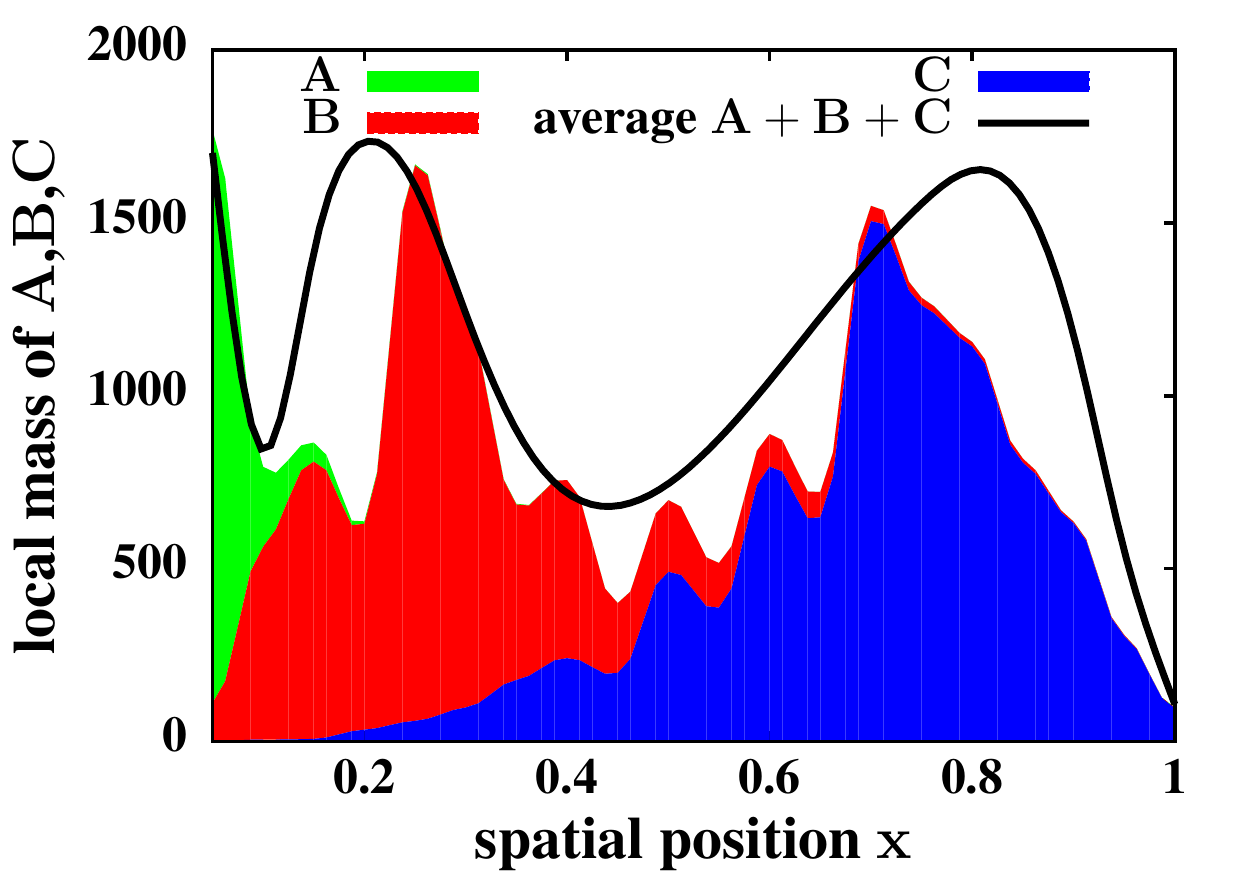}
  }
  \renewcommand{\thesubfigure}{d}
\subfloat[]{\label{fig_agg4}
  \includegraphics[width=7.3cm,height=5.0cm]{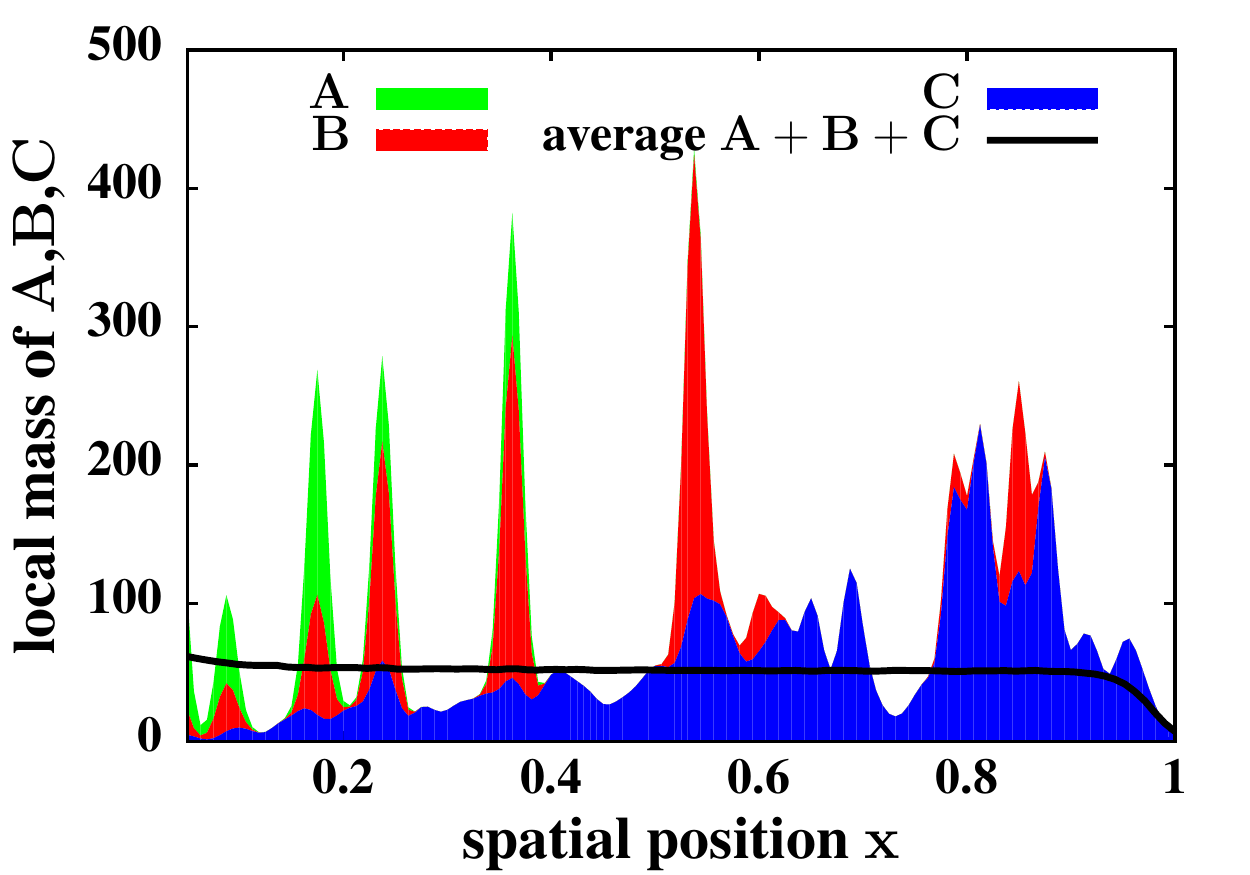}
  }
\caption{Typical steady state configurations in the aggregate representation for (a) pure VT (b) pure CP (c) VT-dominated transport with rare sub-cisternal movement, akin to the Cisternal Progenitor model and 
(d) CP-dominated transport with fission of single C particles. Solid black lines represent time-averaged profiles of the total mass.
All parameters same as in fig. 2 of main paper.
}
  \label{fig_agg}
\end{figure*}

\section{Typical steady state configurations in aggregate representation}
Steady state configurations of mass for various transport models (see fig. 2 of the main paper) can also be shown in an alternative `aggregate representation' where 
 the local mass of $A$, $B$, $C$ at spatial position $x$ is represented by 
 the  heights of the green, red, blue columns at $x$. This representation is particularly useful in tracking
 changes in mass due to perturbations, for instance in movies S9-S12, or for detecting fluctuations of the mass 
 profiles about the time-averaged concentrations (see solid black lines in fig. \ref{fig_agg}, also fig. \ref{fig4}).

\section{Dynamical measurements}
We consider below two kinds of dynamical measurement which address: (i) how the number of tagged particles decays in time after the full system is tagged at $t=0$, and 
(ii) how the compositional entropy  changes during reassembly of the system from an unpolarized state.
 \begin{figure}[h]
\subfloat[] {
\centering
\includegraphics[width=7cm,height=4.8cm]{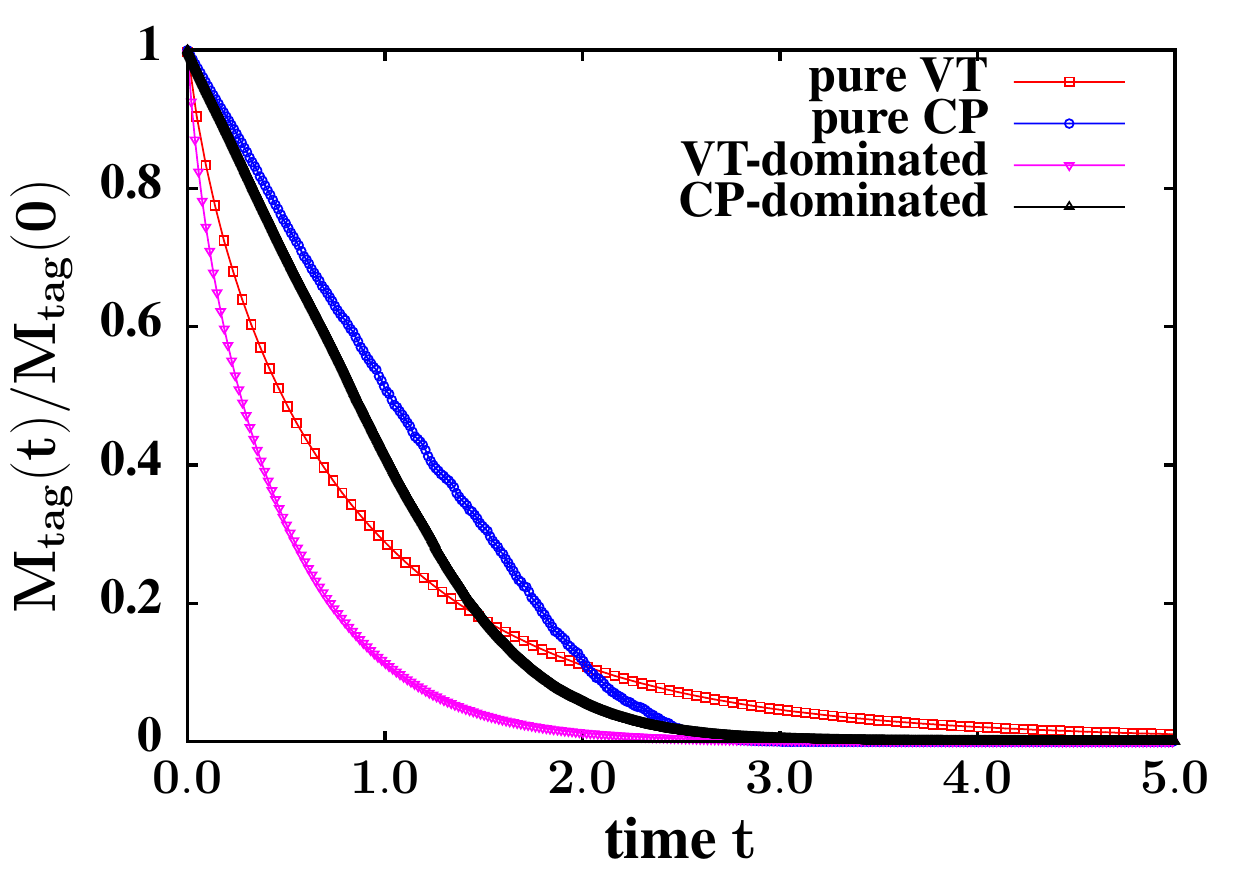}
\label{fig9a}}
\subfloat[]{
\includegraphics[width=7cm,height=4.8cm]{entropy_region}
\label{fig9b}}
\caption{(a) Fraction of tagged particles $M_{tag}(t)/M_{tag}(0)$ remaining in the system at time $t$ vs. $t$ (obtained by averaging over $100$ ensembles). Scale of X axis is 
$1\equiv 10^{5} t.u.$ for pure VT and VT-dominated cases, $1\equiv 10^{4} t.u.$ for the pure CP case and $1\equiv 5\times 10^{3} t.u.$ for the CP-dominated case. 
$M_{tag}(t)/M_{tag}(0)$ decays linearly in the pure CP case but deviates from linear decay in CP-dominated case. Exponential decay for 
pure VT and VT-dominated cases.
(b) Compositional entropy $S_{\Omega}(t)$ vs. $t$ for the region $\Omega:0.5<x<0.75$ during reassembly from an unpolarized state (averaged over $100$ ensembles). Scale of X axis is 
$1\equiv 10^{4} t.u.$. $S_{\Omega}(t)$ decreases during reassembly for pure VT case but changes non-monotonically in the pure CP case. All parameters same as in figs. 2 and 3 of main paper.
}
\label{fig9}
\end{figure}

\subsection{Dynamics of (fluorescently) tagged particles}
\paragraph*{}
We perform measurements inspired by iFRAP (inverse Fluorescence Recovery After Photobleaching) experiments in which the cargo pool in the entire Golgi region is fluorescently tagged and highlighted and then the subsequent decay of fluorescence 
monitored to infer the exit kinetics of the cargo molecules \cite{Patterson1}.  The observation of exponential decay of fluorescence in these experiments has been used to argue against the cisternal progression hypothesis \cite{Patterson1}. To further explicate these
 findings, we implement a similar protocol in our model, by tagging all the particles present in the system at an arbitrary time instant $t=0$ in steady state, and monitoring how the proportion of tagged particles 
 decays in time. Figure \ref{fig9}\subref{fig9a} shows $M_{tag}(t)/M_{tag}(0)$ (averaged over $100$ ensembles) for each of the four cases - pure VT, pure CP, VT-dominated and CP-dominated. A clear linear decay in tagged particle concentration 
 occurs only for the pure CP
 case. Exponential decay is observed for both the pure VT case and the VT-dominated case where cisternal movement provides a secondary track for traffic. Further, even in the CP-dominated case, where the bulk of the transport is through cisternal 
 progression, there is a significant deviation from linear decay. This suggests that the observation of exponential decay in fluorescence experiments is consistent with cisternal progression as long as there is concomitant vesicular movement.
 A similar conclusion is reported in \cite{Dmitrieff} where the iFRAP experiments in \cite{Patterson1} were quantitatively analyzed to show that the exponential decay can be explained by a number of alternative scenarios involving vesicular 
 efflux in addition to cisternal movement.

\subsection{Dynamics of reconstitution after disassembly}  
\paragraph*{}
We consider below the dynamics of re-assembly of the system in the pure VT and pure CP scenarios, and in the process also distinguish reassembly from de novo formation. In our terminology, reassembly refers 
to the reconstitution of the system from an initial state in which the molecules and markers that comprise the cisternae are all present in the Golgi region but distributed in a uniform, unpolarized manner. De novo formation,
on the other hand, refers to the re-formation of structures starting from an initially `empty' state, i.e. with molecules resorbed into the ER and/or dispersed far into the cytoplasm.
\paragraph*{}
To study re-assembly in Monte Carlo simulations, an initial state is constructed  by redistributing all the A,B,C particles (that constitute the steady state structures for a specific set of rates) uniformly across the system. We then 
switch on all transport and interconversion processes at the same rates and monitor how the number of A,B,C particles in different regions evolves in time (see also movies (S7) and (S8)). As in the case of de novo formation, this information 
can be used to compute the region-wise compositional entropy $S_{\Omega}(t)$ (see main paper for a formal definition). As the structures re-assemble, $S_{\Omega}(t)$ decreases with time in the pure VT case and changes non-monotonically in the pure CP cases 
for the region $0.5<x\leq 0.75$ (fig. \ref{fig9}\subref{fig9b}). In general,  $S_{\Omega}(t)$ can show complex non-monotonic behaviour in both limits depending on the region of interest. This is in contrast with de novo formation where the dynamics of the 
region-wise entropy provides a clear way of discriminating between the two mechanisms, increasing with $t$ during formation in the pure CP case, and decreasing with 
$t$ in the pure VT case (see fig. 4(f) of the main paper).
\paragraph*{}
The contrast in the dynamics of $S_{\Omega}(t)$ during de novo formation and reassembly in the pure CP limit highlights the sensitivity of dynamical measurements to the initial state of the system. The difference
between the two dynamics can be rationalized as follows. During de novo formation, early compartments mature into medial and then trans compartments, thus 
creating mixed compartments, and increasing the compositional entropy. Eventually as this process goes on, the creation of mixed compartments due to maturation of early compartments is balanced by the loss of mixed compartments due to 
maturation into late compartments, so that $S_{\Omega}$  reaches a time-independent value. However, in the early stages of de novo formation, when this balance has not yet been achieved, $S_{\Omega}(t)$ increases with time.
In the case of reassembly, on the other hand, early as well as late components are already present in the system in a completely random, unpolarized manner. Reassembly thus essentially involves the re-emergence of polarity in the system,
 and is thus qualitatively different from de novo formation. 

\section{Movies}
\paragraph*{Steady state fluctuations [(S1)-(S4)]\\}
Movies (S1)-(S4) follow the self-organized structures in each of the 4 scenarios (pure VT (S1), VT-dominated (S2), pure CP (S3), CP-dominated (S4)) over a period of time, and provide a window into the dynamics and fluctuations of
these structures in steady state. The structures are represented by separate intensity plots for each of the three species $A,B,C\ldots$, with the intensity of green at any spatial position $x$ being proportional to the abundance of A particles 
at $x$ and so on.\\
{\bf (S1)}: In the pure VT limit, A, B, C particles are stably localized in different region of the system. The abundance of particles in any region shows only minor variation over time.\\
{\bf (S2)}: In the VT-dominated limit, there is greater variation over time, with small chunks breaking off from one structure and fusing with the next (Note the movement of light red plus light blue bars between the B-rich and C-rich 
compartments).\\
{\bf (S3)}: In the pure CP limit, there is a high turnover of cisternae-- any macroscopic region of the system has large mobile cisternae entering and exiting, leading to strong fluctuations in molecular abundances over time. It is also possible to 
visualize the maturation process by following an individual cisterna over time-- a typical cisterna changes color from primarily green near the cis end to green-red to red-blue as it moves, becoming primarily blue near the trans end. \\
{\bf (S4)}: The CP-dominated limit is qualitatively similar to the pure CP limit, except near the trans end where cisternae disintegrate into smaller fragments before leaving the system (Contrast the light blue bars near the trans end in the
CP-dominated limit with the intense blue bars in the pure CP limit).

\paragraph*{De novo formation [(S5)-(S6)]\\}
Movies (S5)-(S6) show how structures form \emph{de novo} in the pure VT limit (S5) and pure CP limit (S6), starting from an initial condition in which the system is completely `empty'. \\
{\bf (S5)}: In the pure VT limit, compartment formation involves a long period of accumulation of A, B and C particles in separate regions of the system, with no mixed compartments forming. A, B, and C compartments regenerate over different
timescales, with the cis (A) compartment taking the shortest and the trans (C) compartment taking the longest time to form. \\
{\bf (S6)}: In the pure CP limit, the system regenerates over a time scale that is  the same as the time required for cisternae to traverse the system (see movie S3). 
Regeneration is characterised by the formation and movement of mixed (two-color) compartments.
\paragraph*{Reassembly dynamics [(S7)-(S8)]\\}
Movies (S7)-(S8) show how cisternae reassemble in the pure VT limit (S7) and pure CP limit (S8), starting from an unpolarized state in which all structures are dissipated and their contents uniformly dispersed across the system. \\
{\bf (S7)}: In the pure VT limit, compartments reassemble through the localization and aggregation of particles of different species (colors) in different regions. \\
{\bf (S8)}: In the pure CP limit, compartment formation involves two stages-- an initial `fast' phase in which all types of particles come together into small mixed aggregates, and a subsequent longer phase in which chemical polarity re-emerges 
as the A-rich structures formed from freshly injected particles traverse the system, maturing into B-rich and then C-rich compartments in different stages of their progression.  
\paragraph*{Response to change in influx [(S9)-(S12)]\\}
Movies (S9)-(S10) show how the mass profiles change in response to a small drop (S9) or rise (S10) in influx in the pure VT limit. (S11)-(S12) show the corresponding response to a small drop (S11) or rise (S12) in the pure CP limit.
To depict changes in mass accurately, we use an alternative representation in which the number of particles
of type A, B, C at a spatial position is proportional to the height of the green, red or blue column respectively at that position.\\
{\bf(S9)-(S10)}: A small change in influx ($10\%$ drop in (S9) and $6.25\%$ rise in (S10)) causes the mass profile to shift significantly from its unperturbed state (solid black line). The proportionate change is higher near the peaks of the mass 
profiles than at the valleys. Different compartments respond over appreciably different time scales, with the cis cisternae attaining their new size much earlier than the trans cisternae. The systematic change in cisternal size is easily distinguishable
from steady state fluctuations.\\
{\bf(S11)-(S12)}: A $20\%$ drop (S11) or rise (S12) in influx causes a modest change in cisternal size. 
Dashed and solid black lines represent the typical cisternal size (obtained by averaging over many ensembles) in the unperturbed state and after a $20\%$  drop/rise in influx respectively. Note how the cisternae in the 
movies can be significantly larger or smaller than this average size making it difficult to distinguish the systematic change in size due to altered influx from usual stochastic fluctuations.
\paragraph*{Response to exit block [(S13)-(S14)]\\}
Movies (S13)-(S14) show how cisternae respond to a blockade of the exit site at the trans end in the pure VT (S13) and pure CP (S14) limits. In both limits, exit blocks induce a piling up of particles at the exit site, with little or no 
change in the rest of the system. In our model, this pile-up continues indefinitely in time, thus pointing towards the need for an additional mechanism that stabilizes the enlarged trans cisterna in the presence of an exit block.

\end{document}